\documentclass[journal]{IEEEtran}
\usepackage{cite,url}
\usepackage{amsmath,amssymb,amsfonts,amstext}
\usepackage{bbm}
\usepackage{epsfig,overpic,psfrag}
\usepackage{graphicx}
\usepackage{algorithm}
\usepackage{algorithmic}
\usepackage{tikz}
\usepackage{epsfig,graphicx,overpic,psfrag}
\DeclareMathOperator{\cov}{cov}
\newtheorem{thm}{Theorem}
\newtheorem{lemma}{Lemma}

\newtheorem{defn}{Definition}
\newtheorem{prop}{Proposition}
\newtheorem{cor}{Corollary}

\begin{document}

\sloppy

\title{On Optimal Fronthaul Compression and Decoding Strategies for Uplink Cloud Radio Access Networks}

\author{Yuhan~Zhou, Yinfei~Xu,~\IEEEmembership{Student Member,~IEEE}, Wei~Yu,~\IEEEmembership{Fellow,~IEEE}, Jun~Chen,~\IEEEmembership{Senior Member,~IEEE}
\thanks{Manuscript received January 19, 2016; revised June 6, 2016; accepted October 2, 2016. This work was supported by the Natural Sciences and Engineering Research Council (NSERC) of Canada. In addition, the work of Y.~Xu was supported in part by a grant from University Grants Committee of the Hong Kong Special Administrative Region, China (Project No. AoE/E-02/08), and by the National Natural Science Foundation of China under Grant 61401086. The material in this paper was presented in part at IEEE International Symposium on Information Theory, Hong Kong, China, June 2015.

Y.~Zhou was with the The Edward S. Rogers Sr. Department of Electrical
and Computer Engineering, University of Toronto, Toronto, ON M5S 3G4
Canada. He is now with Qualcomm Technologies Inc., San Diego, CA 92121
USA (email: yzhou@ece.utoronto.ca).

Y.~Xu was with the School of Information Science and Engineering, Southeast University, Nanjing  210096, China. He is now with
the Department of Information Engineering, Institute of Network Coding, The Chinese University of Hong Kong, Hong Kong (e-mail: yinfeixu@inc.cuhk.edu.hk).

W.~Yu is with The Edward S. Rogers Sr. Department of Electrical and
Computer Engineering, University of Toronto, Toronto, ON M5S 3G4
Canada (e-mail: weiyu@ece.utoronto.ca).

J.~Chen is with the Department of Electrical and Computer
Engineering, McMaster University, Hamilton, ON L8S 4L8, Canada
(e-mail: junchen@ece.mcmaster.ca)

Communicated by O. Simeone, Associate Editor for Communications.

Copyright (c) 2014 IEEE. Personal use of this material is permitted.  However, permission to use this material for any other purposes must be obtained from the IEEE by sending a request to pubs-permissions@ieee.org.

}
}

\markboth{IEEE TRANSACTIONS ON INFORMATION THEORY, ~VOL. 00,~NO. 0, XXX 2016}%
{Shell \MakeLowercase{\textit{et al.}}: Bare Demo of IEEEtran.cls for Journals}
%



\maketitle


\begin{abstract}
This paper investigates the compress-and-forward scheme for an uplink
cloud radio access network (C-RAN) model, where multi-antenna
base-stations (BSs) are connected to a cloud-computing based central
processor (CP) via capacity-limited fronthaul links. The BSs compress
the received signals with Wyner-Ziv coding and send the representation bits to the CP; the
CP performs the decoding of all the users' messages.  Under this
setup, this paper makes progress toward the optimal structure of the
fronthaul compression and CP decoding strategies for the
compress-and-forward scheme in C-RAN. On the CP decoding strategy design, this paper shows that under a sum
fronthaul capacity constraint, a generalized successive decoding
strategy of the quantization and user message codewords that allows
arbitrary interleaved order at the CP achieves the same rate region
as the optimal joint decoding.  Further, it is shown that a practical
strategy of successively decoding the quantization codewords first,
then the user messages, achieves the same maximum sum rate as joint
decoding under individual fronthaul constraints.  On the joint
optimization of user transmission and BS quantization strategies, this
paper shows that if the input distributions are assumed to be
Gaussian, then under joint decoding, the optimal quantization scheme
for maximizing the achievable rate region is Gaussian. Moreover, Gaussian input and Gaussian quantization with joint decoding
achieve to within a constant gap of the capacity region of the
Gaussian multiple-input multiple-output (MIMO) uplink C-RAN model.
Finally, this paper addresses the computational aspect of optimizing
uplink MIMO C-RAN by showing that under fixed Gaussian input, the sum rate maximization problem over the Gaussian quantization
noise covariance matrices can be formulated as convex optimization
problems, thereby facilitating its efficient solution.
\end{abstract}



\section{Introduction}
\label{chapOpt-sec-intro}

Cloud Radio Access Network (C-RAN) is an emerging mobile network
architecture in which base-stations (BSs) in multiple cells are
connected to a cloud-computing based central processor (CP) through
wired/wireless fronthaul links. In the deployment of a C-RAN system,
the BSs degenerate into remote antennas heads implementing only radio
functionalities, such as frequency up/down conversion, sampling,
filtering, and power amplification. The baseband operations at the BSs
are migrated to the CP. The C-RAN model effectively virtualizes
radio-access operations such as the encoding and decoding of user
information and the optimization of radio resources
\cite{Simeone15}. Advanced joint multicell processing
techniques, such as the coordinated multi-point (CoMP) and network
multiple-input multiple-output (MIMO), can be efficiently
supported by the C-RAN architecture, potentially enabling
significantly higher data rates than conventional cellular networks
\cite{Gesbert10}.

\begin{figure} [t]
    \centering
    \begin{overpic}[width=0.46\textwidth]{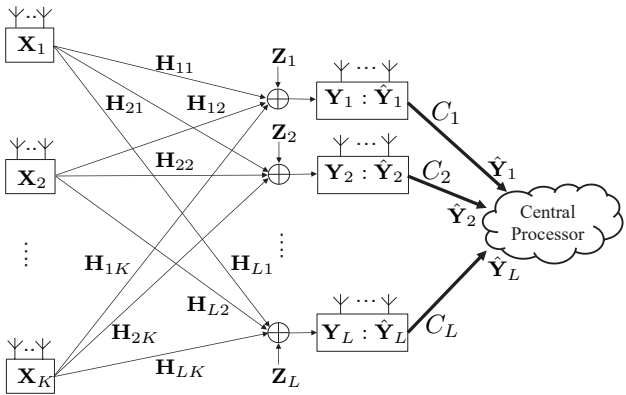}
    \put(52,46.5){\footnotesize $\mathbf{Y}_1:\hat{\mathbf{Y}}_1$}
    \put(52,34.5){\footnotesize $\mathbf{Y}_2:\hat{\mathbf{Y}}_2$}
    \put(51.5,9){\footnotesize $\mathbf{Y}_L:\hat{\mathbf{Y}}_L$}
    \put(3,55){\footnotesize $\mathbf{X}_1$}
    \put(3,33.5){\footnotesize $\mathbf{X}_2$}
    \put(3,2){\footnotesize $\mathbf{X}_K$}
    \put(43.5,53){\footnotesize $\mathbf{Z}_1$}
    \put(43.5,41){\footnotesize $\mathbf{Z}_2$}
    \put(43.5,2){\footnotesize $\mathbf{Z}_L$}
    \put(25,52){\footnotesize $\mathbf{H}_{11}$}
    \put(17,45.5){\footnotesize $\mathbf{H}_{21}$}
    \put(30,45.5){\footnotesize $\mathbf{H}_{12}$}
    \put(13.7,20){\footnotesize $\mathbf{H}_{1K}$}
    \put(37,20){\footnotesize $\mathbf{H}_{L1}$}
    \put(18,9){\footnotesize $\mathbf{H}_{2K}$}
    \put(30,13.5){\footnotesize $\mathbf{H}_{L2}$}
    \put(25,36.5){\footnotesize $\mathbf{H}_{22}$}
    \put(25,3.5){\footnotesize $\mathbf{H}_{LK}$}
    \put(78,35){\footnotesize $\hat{\mathbf{Y}}_1$}
    \put(71.5,27.5){\footnotesize $\hat{\mathbf{Y}}_2$}
    \put(78,19){\footnotesize $\hat{\mathbf{Y}}_L$}
    \put(69,44){$C_1$}
    \put(67.5,34.5){$C_2$}
    \put(68,10){$C_L$}
    \end{overpic}
    \caption{The uplink C-RAN model under finite-capacity fronthaul constraints}
    \label{fig:C-RAN-math}
\end{figure}

This paper considers the uplink of a MIMO C-RAN system under
finite-capacity fronthaul constraints, as shown in
Fig.~\ref{fig:C-RAN-math}, which consists of multiple remote users
sending independent messages to the CP through multiple BSs serving
as relay nodes. Both the user terminals and the BSs are equipped with
multiple antennas. The BSs and the CP are connected via
noiseless fronthaul links with finite capacity. %
This channel model can be thought of as a two-hop relay
network, with an interference channel between the users and the BSs,
followed by a noiseless multiple-access channel between the BSs and
the CP. This paper assumes that a compress-and-forward relaying strategy is employed, 
in which the relaying BSs perform distributed lossy source coding to compress the received signals and forward the
representation bits to the CP through digital fronthaul links, and all the user
messages are eventually decoded at the CP. The lossy source coding implemented at BSs involves Wyner-Ziv coding typically consisting of quantization followed by binning in order to achieve high compression efficiency by leveraging the correlation between the received signals across different BSs, which is different from the point-to-point fronthaul compression implemented in today's conventional C-RAN systems.

A key question in the design of compress-and-forward strategy in
uplink C-RAN is the optimal input coding strategy at the user
terminals, the optimal relaying strategy at the BSs, and the optimal
decoding strategy at the CP. Toward this end, this paper restricts
attention to the strategy of compressing the received signals at the BSs,
then either \emph{joint decoding} of the quantization and message
codewords simultaneously, or \emph{generalized successive decoding} of
the quantization and message codewords in some arbitrary order at the
CP. Under this assumption, this paper makes the following contributions
toward revealing the structure of the optimal compress-and-forward
strategy.

First, motivated by the fact that successive decoding is much easier
to implement than joint decoding, we seek to understand whether
successive decoding at the CP can perform as well as joint decoding.
Toward this end, this paper shows that generalized successive decoding indeed achieves the
same rate region as joint decoding for an uplink C-RAN model under a sum fronthaul
constraint. Further, although not necessarily so for the general rate
region, if one focuses on maximizing the sum rate, the particular strategy of 
successively decoding the quantization codewords first, then the user
messages, achieves the optimal sum rate.

Second, we seek to understand the optimal input distribution and
quantization schemes in uplink C-RAN. Although it is well known that
joint Gaussian strategies are not necessarily optimal, this paper
shows that if we fix the input distribution to be Gaussian, then the
optimal quantization scheme is Gaussian under joint decoding, and vice
versa. Moreover, joint Gaussian signaling can be shown to achieve the
capacity region of the Gaussian multiple-input multiple-output (MIMO)
uplink C-RAN model to within a constant gap.
Finally, this paper makes progress on the computational front by
showing that under the joint Gaussian assumption, the optimization of
the quantization covariance matrices for maximizing the sum rate can be formulated as a convex optimization problem. These results suggest that joint Gaussian input signaling and Gaussian quantization is a
suitable strategy for the uplink C-RAN.

\subsection{Related Work}
\label{subsec:relatework}

The achievable rate region of compress-and-forward with joint decoding
for the uplink C-RAN model was first characterized in~\cite{Sand09}
for a single-transmitter model then in~\cite{Sander09MIMO} for the
multi-transmitter case. However, the number of rate constraints in the
joint decoding rate region grows exponentially with the size
of the network~\cite[Proposition IV.1]{Sand09}, which makes the
evaluation of the achievable rate computationally prohibitive. The
achievable rate region of the compress-and-forward strategy with
practical successive decoding, 
in which the quantization codewords are decoded first, then the user
messages are decoded based on the recovered quantization codewords,
has also been studied for the uplink C-RAN model~\cite[Theorem 1]{Sand08}.
One of the objectives of this paper is to illustrate the relationship
between joint decoding and successive decoding. In the existing
literature, the
equivalence between these two decoding schemes is first demonstrated
for single-source, single-destination, and single-relay
networks~\cite[Appendix 16C]{El2011network}, then shown for
single-source, single-destination, and multiple-relay
networks~\cite{wu2013optimal}, under either block-by-block forward
decoding or block-by-block backward decoding. This paper further
demonstrates that in the case of uplink C-RAN, which is a
multiple-source, single-destination, multiple-relay network, the
optimality of successive decoding still holds under suitable
conditions.

In general, it is challenging to find the optimal joint input and
quantization noise distributions
that maximize the achievable rate of the compress-and-forward scheme
for uplink C-RAN. Gaussian signaling is not necessarily optimal---in
particular, in a simple example of uplink C-RAN with one user and two
BSs shown in~\cite{Sand08}, binary input is shown to outperform Gaussian input for a broad range of signal-to-noise ratios (SNRs).
However, Gaussian input and Gaussian quantization can be shown to be
approximately optimal. In fact, the uplink C-RAN model is an example of a general Gaussian relay network with multiple sources and a
single destination for which a generalization of compress-and-forward
with joint decoding (referred to as noisy network coding
scheme~\cite{Avest11, Lim11,yassaee2011slepian,hou2015short}) and with Gaussian input and Gaussian quantization can be
shown to achieve to within a constant gap to the information
theoretical capacity of the overall network. Instead of using noisy
network coding, our previous work \cite{zhou2014JSAC} shows that
successive decoding can achieve the sum capacity of uplink C-RAN to
within constant gap, if the fronthaul links are subjected to a sum
capacity constraint. In this work, we further demonstrate that the
compress-and-forward scheme with joint decoding can achieve
to within a constant gap to the entire capacity region of the uplink
C-RAN model with individual fronthaul constraints; same is true for successive decoding under suitable
condition.


An important theoretical result obtained in this paper is that if
the input distributions of the uplink C-RAN model are fixed to be
Gaussian, then Gaussian quantizer is in fact optimal under joint
decoding. Finding the optimal quantization for the C-RAN model is
related to the mutual information constraint problem~\cite{Tian09},
for which entropy power inequality is used to show that Gaussian
quantization is optimal for a three-node relay network with Gaussian
input. However, it is challenging to extend this approach to the
uplink C-RAN model, which has multiple sources. This paper provides a
novel proof of the optimality of Gaussian quantization based on the
de Bruijn identity and the Fisher information inequality.
The idea of the proof is inspired by the connection between the C-RAN model and the CEO
problem in source coding \cite{Berger96}, where a source is described
to a central unit by remote agents with noisy observations.
The solution to the CEO problem is known for the scalar Gaussian case~\cite{Oohama05, Prabhakaran04}; significant recent progress has
been made in the vector case, e.g., \cite{Wang14}. The similarity between the uplink C-RAN model and the CEO problem has been noted in \cite{Sand08}, based on which a capacity upper bound for the uplink C-RAN model is established.
In this paper, we use techniques for establishing the outer bound for the
Gaussian vector CEO problem~\cite{EU14} to prove the optimality of
Gaussian quantization. We also remark the connection between this
quantization optimization problem and the information bottleneck
method~\cite{tishby1999information}, for which joint Gaussian
distribution is shown to be Pareto optimal. The technique used in this
paper is a significantly simpler alternative to the enhancement technique given
in~\cite{Liu07}.

This paper also makes progress in observing that the optimization of
Gaussian quantization noise covariance matrices for maximizing the
(weighted) sum rate of uplink C-RAN can be reformulated as a convex
optimization problem. 
The quantization noise covariance optimization problem for uplink C-RAN
has been considered extensively in the literature. Various optimization algorithms have been developed to
maximize the achievable rates of the compress-and-forward scheme for the case of either successive decoding of the quantization codewords followed by the user messages~\cite{Del09, Park12} or joint decoding of the quantization codewords and user messages simultaneously~\cite{park2013jointSPL}.  In particular, a zero-duality gap result has been shown for the weighted sum rate maximization problem under a sum fronthaul capacity constraint in \cite{Del09} based on a time-sharing argument to facilitate the algorithm design for searching optimal quantization noise covariance matrices. However, the optimization problems formulated in these works (i.e., \cite{Del09}, \cite{Park12}, \cite{park2013jointSPL}) are inherently nonconvex, hence only locally convergent algorithms are obtained. Instead, this paper provides a convex formulation of the problem that allows globally optimal
Gaussian quantization noise covariance matrices to be found.  Note that here the optimization of the quantization
noise covariance matrix is performed under the fixed Gaussian input.
The joint optimization of the input
signal and quantization noise covariance matrices remains a computationally challenging difficult
problem~\cite{zhou2015SP}.

\subsection{Main Contributions}

This paper establishes several information theoretic results on the compress-and-forward scheme for
the uplink MIMO C-RAN model with finite-capacity fronthaul links.
A summary of our main contributions is as follows:
\begin{itemize}
\item This paper demonstrates that generalized successive decoding
for compress-and-forward, which allows the decoding of the quantization
and user message codewords in an arbitrary order, can achieve the
same rate region as joint decoding for compress-and-forward under a sum fronthaul capacity
constraint.  Further, successive decoding of the quantization
codewords first, then the user message codewords, can achieve the
same maximum sum rate as joint decoding under individual fronthaul
constraints.
\item This paper shows that under Gaussian input and Gaussian quantization,
compress-and-forward with joint decoding achieves to within a constant gap of the capacity region of the uplink
MIMO C-RAN model. Combining with the result
above, the same constant-gap result also holds for generalized
successive decoding under a sum fronthaul constraint and for successive
decoding for sum rate maximization.
\item This paper shows that under fixed Gaussian input, Gaussian
quantization maximizes the achievable rate region under joint
decoding.  Combining with the optimality result for successive
decoding, this also implies that under fixed Gaussian input,
Gaussian quantization is optimal for generalized successive decoding
under a sum fronthaul constraint, and for successive decoding for sum
rate maximization.
\item Under joint Gaussian signaling and Gaussian quantization,
the optimization of quantization noise covariance matrices for
maximizing weighted sum rate under joint decoding and for
maximizing sum rate under successive decoding can be
formulated as convex optimization problems, which facilitate their
efficient solution.
\end{itemize}

\subsection{Paper Organization and Notation}
\label{subsec:organiz-notation}
The rest of the paper is organized as follows.
Section~\ref{chapOpt-sec-Prelim} introduces the channel model for
the uplink MIMO C-RAN and characterizes the achievable rate regions
for compress-and-forward schemes with joint decoding and generalized
successive decoding respectively.  Section~\ref{chapOpt-sec-OptSD}
demonstrates the rate-region optimality of generalized successive
decoding under a sum fronthaul constraint and the sum-rate optimality
of successive decoding.  Section~\ref{chapOpt-sec-OptimalQ} focuses on
establishing the optimality of Gaussian quantizers with joint decoding
under Gaussian input. In addition, Section~\ref{chapOpt-sec-OptimalQ}
also establishes the approximate capacity of the uplink MIMO C-RAN
model to within constant gap, and shows the convex formulation of the
(weighted) sum rate maximization problems over the quantization noise
covariance matrices.
Section~\ref{chapOpt-sec-conclustion} concludes the paper.

Notation: Boldface letters denote vectors or matrices, where context should make the distinction clear. Superscripts $(\cdot)^{\mathsf{T}}$, $(\cdot)^{\dag}$ and $(\cdot)^{-1}$ denote transpose operation, Hermitian transpose and matrix inverse operators; $\mathbb{E}[\cdot]$ and $\mathrm{Tr}(\cdot)$ denote expectation and matrix trace operators; 
$\textrm{co}(\cdot)$ denotes the convex closure operation; $p(\cdot)$ denotes the probability distribution function in this paper. We use $\mathbf{X}^j_i = \left(\mathbf{X}_i, \mathbf{X}_{i+1}, \ldots, \mathbf{X}_j\right)$ to denote a  matrix with $(j-i+1)$ columns for $1\leq i \leq j$. For a vector/matrix $\mathbf{X}$, $\mathbf{X}_{\mathcal{S}}$ denotes a vector/matrix with elements whose indices are elements of $\mathcal{S}$. Given matrices $\{\mathbf{X}_1,\ldots, \mathbf{X}_L\}$, $\mathrm{diag}\left(\{\mathbf{X}_{\ell}\}_{\ell=1}^L\right)$ denotes the block diagonal matrix formed with $\mathbf{X}_{\ell}$ on the diagonal. For random vectors $\mathbf{X}$ and $\mathbf{Y}$, $\mathbf{J}(\mathbf{X}|\mathbf{Y})$ denotes the Fisher information matrix of $\mathbf{X}$ conditional on $\mathbf{Y}$; $\textrm{cov}(\mathbf{X}|\mathbf{Y})$ denotes the covariance matrix of $\mathbf{X}$ conditional on $\mathbf{Y}$.

\section{Achievable Rate Regions for Uplink C-RAN}
\label{chapOpt-sec-Prelim}

\subsection{Channel Model}
\label{subsec:channel}
This paper considers an uplink C-RAN model, where $K$ mobile users communicate with
a CP through $L$ BSs, as shown in Fig.~\ref{fig:C-RAN-math}. The noiseless digital fronthaul
link connecting the BS $\ell$ to the CP has the capacity of $C_{\ell}$ bits per complex dimension. The fronthaul capacity $C_{\ell}$ is the maximum long-term average throughput of the $\ell$th fronthaul link, i.e.,
$\lim\limits_{n\rightarrow \infty}\frac{1}{n}\sum_{i=1}^n C_{\ell}(i)\leq C_{\ell}$,
where $C_{\ell}(i)$ represents the instantaneous transmission rate of the $\ell$th fronthaul link at the $i$th time slot. Each user terminal is equipped with $M$ antennas; each BS is equipped with $N$ antennas. Perfect channel state information (CSI) is assumed to be available to all the BSs and to the CP. For simple notation, we denote $\mathcal{K}=\{1,\cdots,K\}$ and $\mathcal{L}=\{1,\cdots,L\}$ in this paper.

Let $\mathbf{X}_k \in \mathbb{C}^M$ be the signal transmitted by the $k$th user, which is subject to per-user transmit power constraint of $P_k$, i.e. $\mathbb{E}\left[\mathbf{X}_k\mathbf{X}_k^{\dagger}\right] \leq P_k$. The signal received at the $\ell$th BS can be expressed as
\begin{equation}\label{eqn:channel-model}
\mathbf{Y}_{\ell}=\sum_{k=1}^K\mathbf{H}_{\ell,k}\mathbf{X}_{k}+\mathbf{Z}_{\ell}, \quad \ell=1,2,\ldots, L,
\end{equation}
where $\mathbf{Z}_{\ell}\sim\mathcal{CN}(\mathbf{0},\mathbf{\Sigma}_{\ell})$ represents the additive Gaussian noise
for BS $\ell$ and is independent across different BSs, and $\mathbf{H}_{\ell,k}$ denotes the complex channel matrix
from user $k$ to BS $\ell$.

We consider the compress-and-forward scheme~\cite{cover1979capacity, El2006bounds} applied to the uplink C-RAN
system, in which the BSs compress the received signals $\mathbf{Y}_{\ell}$, and forward the quantization bits to the CP for decoding. At the CP, the user messages are decoded using either joint decoding or some form of successive decoding. In joint decoding, the quantization codewords and the message codewords are decoded \emph{simultaneously}, whereas, in successive decoding, the quantization codewords and messages are decoded \emph{successively} in some prescribed order.
Different orderings can potentially result in different achievable rates.

\subsection{Achievable Rates for Joint Decoding, Successive Decoding, and Generalized Successive Decoding}
\label{subsec:rate-JDvsSD}
In the following, we present the achievable rate region of compress-and-forward with joint decoding and different forms of successive decoding.

\begin{prop}[{\cite[Proposition IV.1]{Sand09}}]
For the uplink C-RAN model shown in Fig.~\ref{fig:C-RAN-math}, the achievable rate-fronthaul region of compress-and-forward with joint decoding, $\mathcal{P}^*_{JD}$, is the closure of the convex hull of all $(R_1,\cdots,R_K, C_1, \ldots, C_L) \in \mathbb{R}_+^{K+L}$ satisfying
\begin{equation}\label{eqn:Rate-JD}
\sum_{k\in \mathcal{T}} R_k < \sum\limits_{\ell\in \mathcal{S}}\left[C_{\ell} - I\left(\mathbf{Y}_{\ell};\hat{\mathbf{Y}}_{\ell}|\mathbf{X}_{\mathcal{K}}\right) \right]
 + I\left(\mathbf{X}_{\mathcal{T}}; \hat{\mathbf{Y}}_{\mathcal{S}^c}|\mathbf{X}_{\mathcal{T}^c}\right)
\end{equation}
for all $\mathcal{T}\subseteq\mathcal{K}$ and $\mathcal{S}\subseteq\mathcal{L}$, for some product distribution $\prod_{k=1}^Kp(\mathbf{x}_k)\prod_{\ell=1}^Lp(\hat{\mathbf{y}}_{\ell}|\mathbf{y}_{\ell})$ such that $\mathbb{E}\left[\mathbf{X}_k\mathbf{X}_k^{\dagger}\right] \leq P_k$ for $k=1,\ldots,K$.
\end{prop}

Note that for the uplink C-RAN model, the rate region (\ref{eqn:Rate-JD}) given by compress-and-forward with joint decoding is identical to the rate region of the noisy network coding scheme~\cite{Lim11}, which is an extension of the compress-and-forward scheme to the general multiple access relay network by using joint decoding at the receiver and block Markov coding at the transmitters.

As a more practical decoding strategy, successive decoding of quantization codewords first, and then the user messages at the CP can also be used in uplink C-RAN. The following proposition states the rate-fronthaul region achieved by successive decoding.

\begin{prop}[{\cite[Theorem 1]{Sand08}}]
For the uplink C-RAN model shown in Fig.~\ref{fig:C-RAN-math}, the achievable rate-fronthaul region of compress-and-forward with successive decoding, $\mathcal{P}^*_{SD}$, is the closure of the convex hull of all $(R_1,\cdots,R_K, C_1, \ldots, C_L)\in \mathbb{R}_+^{K+L}$ satisfying
\begin{equation}
\label{eqn:SD-rate-I}
\sum_{k\in \mathcal{T}} R_k < I\left(\mathbf{X}_{\mathcal{T}}; \hat{\mathbf{Y}}_{\mathcal{L}}| \mathbf{X}_{\mathcal{T}^c}\right), \quad \forall \; \mathcal{T} \subseteq \mathcal{K}, \\
\end{equation}
and
\begin{equation}
\label{eqn:SD-q-I}
I\left(\mathbf{Y}_{\mathcal{S}}; \hat{\mathbf{Y}}_{\mathcal{S}}| \hat{\mathbf{Y}}_{\mathcal{S}^c}\right) < \sum_{\ell\in \mathcal{S}}C_{\ell}, \quad \forall \; \mathcal{S} \subseteq \mathcal{L}.
\end{equation}
for some product distribution $\prod_{k=1}^Kp\left(\mathbf{x}_k\right)\prod_{\ell=1}^L p(\hat{\mathbf{y}}_{\ell}|\mathbf{y}_{\ell})$ such that $\mathbb{E}\left[\mathbf{X}_k\mathbf{X}_k^{\dagger}\right] \leq P_k$ for $k=1,\ldots,K$.
\end{prop}

Note that (\ref{eqn:SD-rate-I}) is the multiple-access rate region, (\ref{eqn:SD-q-I}) represents the Berger-Tung rate region for distributed lossy compression~\cite[Theorem 12.1]{El2011network}, while (\ref{eqn:Rate-JD}) incorporates the joint decoding of the quantization codewords and the user messages. Because of its lower decoding complexity, successive decoding is usually preferred for practical implementation of the uplink C-RAN systems~\cite{Del09, Park12}. Note that in the above strategy, successive decoding applies only to the vector $\mathbf{X}_k$ (user message codewords) and vector $\mathbf{Y}_{\ell}$ (quantization codewords); the elements within vectors $\mathbf{X}_k$ and $\mathbf{Y}_{\ell}$ are still decoded jointly.

It is possible to improve upon the successive decoding scheme by allowing arbitrary interleaved decoding orders between quantization codewords and user message codewords. We call this the generalized successive decoding scheme in this paper. The generalized successive decoding scheme is first suggested in~\cite{zhoulei2013} under the name of joint base-station successive interference cancelation scheme. In such a successive decoding strategy, the set of potential decoding orders includes all the permutations of quantization and user message codewords.

Denote $\pi$ as a permutation on the set of quantization and user message codewords $\left(\hat{\mathbf{Y}}_1,  \hat{\mathbf{Y}}_2, \ldots, \hat{\mathbf{Y}}_L, \mathbf{X}_1,\mathbf{X}_2, \ldots \mathbf{X}_K\right)$. For a given permutation $\pi$, the decoding order is given by the index of the elements in $\pi$, i.e., $\pi(1)\rightarrow \pi(2)\rightarrow  \cdots \rightarrow \pi(L+K)$. For example, consider an uplink C-RAN model as shown in Fig.~\ref{fig:C-RAN-math} with $2$ BSs and $2$ users. If $\pi = \left(\hat{\mathbf{Y}}_{1}, \mathbf{X}_1, \hat{\mathbf{Y}}_{2}, \mathbf{X}_2\right)$, then the decoding of $\hat{\mathbf{Y}}_{2}$ and $\mathbf{X}_2$ can use both previously decoded user messages and quantization codewords as side information. The resulting rate region is characterized as
\begin{equation}
\label{eqn:Rate-GSD-2user}
\begin{cases}
\enspace  R_{1}< I\left(\mathbf{X}_{1}; \hat{\mathbf{Y}}_{1}\right),   \\
\enspace  R_{2}< I\left(\mathbf{X}_{2}; \hat{\mathbf{Y}}_{1}, \hat{\mathbf{Y}}_{2}| \mathbf{X}_{1}\right),
\end{cases}
\end{equation}
for some product distribution $p(\mathbf{x}_1)p(\mathbf{x}_2)p(\hat{\mathbf{y}}_{1}|\mathbf{y}_{1})p(\hat{\mathbf{y}}_{2}|\mathbf{y}_{2})$ that satisfies
\begin{equation}\label{eqn:GSD-q-2user}
\begin{cases}
\enspace C_{1} > I\left(\mathbf{Y}_{1}; \hat{\mathbf{Y}}_{1}\right),  \\
\enspace C_{2} > I\left(\mathbf{Y}_{2}; \hat{\mathbf{Y}}_{2}|\hat{\mathbf{Y}}_{1},\mathbf{X}_{1}\right).
\end{cases}
\end{equation}
Let $\mathcal{I}_{\mathbf{X}_k}$, $\mathcal{I}_{\mathbf{Y}_{\ell}}$ denote the indices of user messages that are decoded before $\mathbf{X}_k$ and $\mathbf{Y}_{\ell}$ under the permutation $\pi$, respectively. Likewise, let $\mathcal{J}_{\mathbf{X}_k}$, $\mathcal{J}_{\mathbf{Y}_{\ell}}$ denote the indices of quantization codewords that are decoded before $\mathbf{X}_k$ and $\mathbf{Y}_{\ell}$ under the permutation $\pi$, respectively. The rate-fronthaul region of generalized successive decoding for uplink C-RAN is stated in the following proposition.

\begin{prop}
For the uplink C-RAN model shown in Fig.~\ref{fig:C-RAN-math}, the achievable rate-fronthaul region of generalized successive decoding with decoding order $\pi$, $\mathcal{P}_{GSD}(\pi)$, is the closure of the convex hull of all $(R_1,\cdots,R_K, C_1, \ldots, C_L) \in \mathbb{R}_+^{K+L}$ satisfying
\begin{equation}
\label{eqn:Rate-GSD}
R_{k} < I\left(\mathbf{X}_{k}; \hat{\mathbf{Y}}_{\mathcal{J}_{\mathbf{X}_k}}|\mathbf{X}_{\mathcal{I}_{\mathbf{X}_k}}\right),  \quad \forall \; k\in \mathcal{K},
\end{equation}
and
\begin{equation}
\label{eqn:GSD-q-I}
C_{\ell} > I\left(\mathbf{Y}_{\ell}; \hat{\mathbf{Y}}_{\ell}|\hat{\mathbf{Y}}_{\mathcal{J}_{\mathbf{Y}_{\ell}}},\mathbf{X}_{\mathcal{I}_{\mathbf{Y}_{\ell}}}\right), \quad \forall \;\ell\in \mathcal{L}.
\end{equation}
for some product distribution $\prod_{k=1}^Kp\left(\mathbf{x}_k\right)\prod_{\ell=1}^Lp(\hat{\mathbf{y}}_{\ell}|\mathbf{y}_{\ell})$ such that $\mathbb{E}\left[\mathbf{X}_k\mathbf{X}_k^{\dagger}\right] \leq P_k$ for $k=1,\ldots,K$.
The generalized successive decoding region $\mathcal{P}^*_{GSD}$  is defined to be the closure of the convex hull of the union of regions $\mathcal{P}_{GSD}(\pi)$ over all possible permutation $\pi$'s,
i.e.,
\begin{equation}
\label{eqn:def-GSD}
\mathcal{P}^*_{GSD}= \mathrm{co}\left(\bigcup\limits_{\pi}\mathcal{P}_{GSD}(\pi)\right).
\end{equation}
\end{prop}

\section{Optimality of Successive Decoding}
\label{chapOpt-sec-OptSD}

In general, we have $\mathcal{P}^*_{SD} \subseteq \mathcal{P}^*_{GSD} \subseteq \mathcal{P}^*_{JD}$. However, successive decoding is more desirable than joint decoding, not only because of its lower complexity, but also due to the fact that its rate region can be more easily evaluated. Thus, there is a tradeoff between complexity and performance in designing decoding strategies for uplink C-RAN. To further understand this tradeoff, this section establishes that: 1) By allowing arbitrary decoding orders of quantization and message codewords, the generalized successive decoding actually achieves the same rate region as joint decoding under a sum fronthaul constraint; 2) The practical successive decoding strategy in which the BSs decode the quantization codewords first, then the user messages, actually achieves the same maximum sum rate as joint decoding under individual fronthaul constraints.

\subsection{Optimality of Generalized Successive Decoding under a Sum Fronthaul Constraint}
\label{subsec:opt-GSD}

This section shows that in the special case where the fronthaul links are subject to a sum capacity constraint,
generalized successive decoding achieves the rate region as joint decoding. In this model,
the fronthaul capacities are constrained by $\sum_{\ell=1}^L C_{\ell} \leq C$ and $C_{\ell}\geq 0$, justifiable in situations where the fronthaul are implemented in shared medium (e.g. wireless fronthaul links), as has been considered in \cite{Del09,zhou2014JSAC}. Under the sum fronthaul capacity constraint $C$, the rate regions achieved by with joint decoding $\mathcal{R}^*_{JD,s}$ is defined as
\begin{multline}
\label{def:Region-JD-SumFront}
\hspace{-3mm}\mathcal{R}^*_{JD,s} = \\
\left\{(R_1,\ldots,R_K)\left|
\begin{array}{c}
(R_1,\cdots,R_K, C_1, \ldots, C_L) \in \mathcal{P}^*_{JD},  \\
\sum_{\ell=1}^L C_{\ell} \leq C, \enspace C_{\ell} \geq 0
\end{array}\right.
\right\}.
\end{multline}
Likewise, the rate region achieved with generalized successive decoding $\mathcal{R}^*_{GSD,s}$ is given by
\begin{multline}
\label{def:Region-GSD-SumFront}
\hspace{-3mm}\mathcal{R}^*_{GSD,s} = \\
\left\{(R_1,\ldots,R_K)\left|
\begin{array}{c}
(R_1,\cdots,R_K, C_1, \ldots, C_L) \in \mathcal{P}^*_{GSD},\\
\sum_{\ell=1}^L C_{\ell}\leq C, \enspace C_{\ell} \geq 0
\end{array}\right.
\right\}.
\end{multline}
The following theorem states the main result of this section.

\begin{thm}\label{thm:GSD=JD-SumFront}
For the uplink C-RAN model with the sum fronthaul capacity constraint $\sum_{\ell=1}^L C_{\ell} \leq C$ and $C_{\ell}\geq 0$, the rate region achieved by generalized successive decoding and joint coding are identical, i.e., $\mathcal{R}^*_{GSD, s}= \mathcal{R}^*_{JD, s}$.
\end{thm}

\begin{IEEEproof}
See Appendix~\ref{append:proof-GSD=JD}.
\end{IEEEproof}

The roadmap for the proof of Theorem~\ref{thm:GSD=JD-SumFront} shares the same idea as the characterization of the rate distortion region for the CEO problem under logarithmic loss~\cite{Courtade14} and the capacity region for the multiple-access channel~\cite{tse1998multiaccess}, which uses the properties of submodular polyhedron (see Appendix~\ref{append:submodular}). Specifically, in order to show $\mathcal{R}^*_{GSD, s}= \mathcal{R}^*_{JD, s}$, we show that under fixed product distribution $\prod_{k=1}^Kp(\mathbf{x}_k)\prod_{\ell=1}^Lp(\hat{\mathbf{y}}_{\ell}|\mathbf{y}_{\ell})$, every extreme point of the polyhedron $(\mathcal{R}^*_{JD, s},C)$ is dominated by the points in the polyhedron defined by $(\mathcal{R}^*_{GSD, s},C)$. We conjecture that Theorem~\ref{thm:GSD=JD-SumFront} holds also for the case of individual fronthaul capacity constraints. However, in that case, finding the dominant faces of  polyhedron $\mathcal{P}^*_{JD}$ becomes much more difficult, it appears non-trivial to extend the current proof to the case of individual fronthaul constraints.

\subsection{Optimality of Successive Decoding for Maximizing Sum Rate}
\label{subsec:opt-SD-SumRate}


As a special instance of generalized successive decoding, successive decoding reconstructs quantization codewords first, then user message codewords in a sequential order.
In what follows, we show that the optimal sum rate achieved by this special successive decoding is the same as that achieved by joint decoding.

Under fixed input distribution and fixed fronthaul capacities $C_{\ell}$, for $\ell=1,\ldots,L$, the maximum sum rate achieved by joint decoding $R^*_{JD,SUM}$ is defined as
\begin{equation}
\label{def:SumRate-JD}
R^*_{JD,SUM}  = \left\{
\begin{array}{cl}
\max & \sum\limits_{k=1}^K R_k  \\
\mathrm{s.t.}  & (R_1,\cdots,R_K, C_1, \ldots, C_L) \in \mathcal{P}^*_{JD}.
\end{array}\right.
\end{equation}
Likewise, the maximum sum rate for successive decoding $R_{SD,SUM}$ is given by
\begin{equation}
\label{def:SumRate-SD}
R^*_{SD,SUM}  = \left\{
\begin{array}{cl}
\max & \sum\limits_{k=1}^K R_k  \\
\mathrm{s.t.} & (R_1,\cdots,R_K, C_1, \ldots, C_L) \in \mathcal{P}^*_{SD}.
\end{array}\right.
\end{equation}
The following theorem demonstrates the optimality of successive decoding for maximizing uplink C-RAN under individual fronthaul constraints.


\begin{thm}\label{thm:opt-VMAC}
For the uplink C-RAN model with fronthaul capacities $C_{\ell}$ shown in Fig.~\ref{fig:C-RAN-math}, the maximum sum rates achieved by successive decoding and joint decoding are the same, i.e., $R^*_{SD, SUM}= R^*_{JD, SUM}$.
\end{thm}

\begin{IEEEproof}
See Appendix~\ref{append:Opt-VMAC}.
\end{IEEEproof}

We remark that Theorem~\ref{thm:opt-VMAC} can be thought as a generalization of a result in~\cite{wu2013optimal} that shows
under block-by-block forward decoding, the compress-and-forward scheme with compression-message successive decoding achieves the same maximum rate as that with compression-message joint decoding for a single-source, single-destination relay network. The uplink C-RAN is a multiple-source, single-destination relay network. If all the user terminals are regarded as one super transmitter, then
it follows from~\cite{wu2013optimal} that successive decoding and joint decoding achieve the same maximum sum rate.
However, the proof in \cite{wu2013optimal} is quite complicated. In this paper, we provide an alternative proof technique for showing the optimality of successive decoding for sum rate maximization in uplink C-RAN. The new proof utilizes the properties of submodular optimization, which is simpler than the proof provided in~\cite{wu2013optimal}. The proofs of Theorem~\ref{thm:opt-VMAC} and Theorem~\ref{thm:GSD=JD-SumFront} illustrate the usefulness of submodular optimization in establishing this type of results.

It is remarked that successive decoding and joint decoding  achieve the same sum rate, but do not achieve the same rate region. The achievable rate region of  generalized successive decoding is in general larger than that of successive decoding. For example, consider the compress-and-forward scheme for maximizing the rate of user $1$, $R_1$, only. The optimal decoding order should be $\mathbf{X}_{\mathcal{K}\setminus \{1\}} \rightarrow \hat{\mathbf{Y}}_{\mathcal{L}} \rightarrow \mathbf{X}_1$. With this decoding order, user 1 can achieve larger rate than using the decoding order of $ \hat{\mathbf{Y}}_{\mathcal{L}} \rightarrow \mathbf{X}_{\mathcal{K}}$, because the decoded user messages $\mathbf{X}_2, \mathbf{X}_3, \ldots, \mathbf{X}_K$ can serve as side information for the decoding of $\hat{\mathbf{Y}}_{\mathcal{L}}$. In general, to maximize a weighted sum rate, one needs to maximize over $(L+K)!$ orderings for generalized successive decoding. The main result of this section shows however that for maximizing the sum rate in uplink C-RAN, successive decoding of the quantization codewords first, and then the user messages is optimal; this reduces the search space considerably to $L!K!$ decoding orders.

\section{Uplink C-RAN with Gaussian Input and Gaussian Quantization}
\label{chapOpt-sec-OptimalQ}

In this section, we specialize to the compress-and-forward scheme for uplink C-RAN with Gaussian input signal at the users and Gaussian quantization at the BSs. Although it is known that joint Gaussian distribution is suboptimal for uplink C-RAN~\cite{Sand08}, Gaussian input is desirable, because it leads to achievable rate regions that can be easily evaluated. In the following section, it is shown that with Gaussian input and Gaussian quantization, compress-and-forward with joint decoding can achieve the capacity region of uplink C-RAN to within a constant gap. The gap depends on the network size but is independent of the channel gain matrix and the SNR. We further establish the optimality of Gaussian compression at the relaying BSs for joint decoding, if the input is Gaussian. These results can be further extended to generalized successive decoding under a sum fronthaul constraint and successive decoding for the maximum sum rate. Additionally, under Gaussian signaling, the optimization of quantization noise covariance matrices for weighted sum-rate maximization under joint decoding and for sum rate maximization under practical successive decoding can be cast as convex optimization problems, thereby facilitating their efficient numerical solution. Throughout this section, we focus on the achievable rates under the fixed Gaussian input, and the fixed fronthaul capacity constraints $C_{\ell}$ for $\ell=1,\ldots,L$.

\subsection{Achievable Rate Regions under Gaussian Input and Gaussian Quantization}
\label{subsec:rate-Gaussian}
We let the input distribution be Gaussian, i.e., $\mathbf{X}_k\sim\mathcal{CN}(\mathbf{0},\mathbf{K}_{k})$, then evaluate the rate regions for the compress-and-forward scheme with joint decoding and successive decoding under
Gaussian quantization, denoted as $\mathcal{R}^G_{JD, GIn}$ and $\mathcal{R}^G_{SD, GIn}$,
respectively. Set $\prod_{\ell=1}^Lp(\hat{\mathbf{y}}_{\ell}|\mathbf{y}_{\ell}) \sim
\mathcal{CN}(\mathbf{y}_{\ell},\mathbf{Q}_{\ell})$, where $\mathbf{Q}_{\ell}$ is the Gaussian
quantization noise covariance matrix at the $\ell$th BS.

With Gaussian input and Gaussian quantization, we have
\begin{equation}
\label{eqn:y_yhat_term}
I(\mathbf{Y}_{\ell};\hat{\mathbf{Y}}_{\ell}|\mathbf{X}_{\mathcal{K}})
	= \log \frac{\left|\mathbf{\Sigma}_{\ell} + \mathbf{Q}_{\ell}\right|}{\left|\mathbf{Q}_{\ell}\right|}
\end{equation}
and  
\begin{align}
\label{eqn:x_yhat_term}
& I\left(\mathbf{X}_{\mathcal{T}};
\hat{\mathbf{Y}}_{\mathcal{S}^c}|\mathbf{X}_{\mathcal{T}^c}\right)  \nonumber \\
& = \log\frac{\left|
\mathbf{H}_{\mathcal{S}^c,\mathcal{T}} \mathbf{K}_{\mathcal{T}}
\mathbf{H}^{\dagger}_{\mathcal{S}^c,\mathcal{T}}  +
\mathrm{diag}\left(\{\mathbf{\Sigma}_{\ell} +
\mathbf{Q}_{\ell}\}_{\ell \in \mathcal{S}^c}\right)\right|}{\left|\mathrm{diag}\left(\{\mathbf{\Sigma}_{\ell} +
\mathbf{Q}_{\ell}\}_{\ell \in \mathcal{S}^c}\right)\right|}.
\end{align}
The achievable rate region (\ref{eqn:Rate-JD}) for joint decoding can be evaluated as
\begin{align}\label{eqn:GaussRateRegion-JD-Q}
\sum_{k\in \mathcal{T}} R_k  & <  \sum_{\ell\in \mathcal{S}}\left[C_{\ell} - \log \frac{\left|\mathbf{\Sigma}_{\ell} + \mathbf{Q}_{\ell}\right|}{\left|\mathbf{Q}_{\ell}\right|} \right]   \nonumber\\
& + \log\frac{\left|
\mathbf{H}_{\mathcal{S}^c,\mathcal{T}} \mathbf{K}_{\mathcal{T}}
\mathbf{H}^{\dagger}_{\mathcal{S}^c,\mathcal{T}}  +
\mathrm{diag}\left(\{\mathbf{\Sigma}_{\ell} +
\mathbf{Q}_{\ell}\}_{\ell \in \mathcal{S}^c}\right)\right|}{\left|\mathrm{diag}\left(\{\mathbf{\Sigma}_{\ell} +
\mathbf{Q}_{\ell}\}_{\ell \in \mathcal{S}^c}\right)\right|},
\end{align}
for all $\mathcal{T}\subseteq\mathcal{K}$ and $\mathcal{S}\subseteq\mathcal{L}$.
%
%
%

Likewise the achievable rate expression (\ref{eqn:SD-rate-I}) for successive decoding becomes
\begin{equation}
\label{eqn:SD-rate-Q}
\sum_{k\in \mathcal{T}} R_k <
\log\frac{\left|
\mathbf{H}_{\mathcal{S}^c,\mathcal{K}} \mathbf{K}_{\mathcal{K}}
\mathbf{H}^{\dagger}_{\mathcal{L},\mathcal{K}}  +
\mathrm{diag}\left(\{\mathbf{\Sigma}_{\ell} +
\mathbf{Q}_{\ell}\}_{\ell \in \mathcal{L}}\right)\right|}{\left|\mathrm{diag}\left(\{\mathbf{\Sigma}_{\ell} +
\mathbf{Q}_{\ell}\}_{\ell \in \mathcal{L}}\right)\right|}, 
\end{equation}
for all $\mathcal{T} \subseteq \mathcal{K}$.

In deriving the fronthaul constraint (\ref{eqn:SD-q-I}), we start with evaluating the mutual
information  
\begin{align}\label{eqn:multual_infor}
& I\left(\mathbf{Y}_{\mathcal{S}}; \hat{\mathbf{Y}}_{\mathcal{S}}| \hat{\mathbf{Y}}_{\mathcal{S}^c}\right) \nonumber \\
& =  I\left(\mathbf{X}_{\mathcal{K}},\mathbf{Y}_{\mathcal{S}}; \hat{\mathbf{Y}}_{\mathcal{S}}| \hat{\mathbf{Y}}_{\mathcal{S}^c}\right)
- I\left(\mathbf{X}_{\mathcal{K}}; \hat{\mathbf{Y}}_{\mathcal{S}}| \mathbf{Y}_{\mathcal{S}}, \hat{\mathbf{Y}}_{\mathcal{S}^c}\right) \nonumber\\
& =  I\left(\mathbf{X}_{\mathcal{K}}; \hat{\mathbf{Y}}_{\mathcal{S}}|\hat{\mathbf{Y}}_{\mathcal{S}^c}\right) + I\left(\mathbf{Y}_{\mathcal{S}};\hat{\mathbf{Y}}_{\mathcal{S}}|\mathbf{X}_{\mathcal{K}},\hat{\mathbf{Y}}_{\mathcal{S}^c} \right) \nonumber \\
& \quad - I\left(\mathbf{X}_{\mathcal{K}}; \hat{\mathbf{Y}}_{\mathcal{S}}| \mathbf{Y}_{\mathcal{S}}, \hat{\mathbf{Y}}_{\mathcal{S}^c}\right) \nonumber\\
& \overset{(a)}{=} I\left(\mathbf{X}_{\mathcal{K}}; \hat{\mathbf{Y}}_{\mathcal{S}}|
\hat{\mathbf{Y}}_{\mathcal{S}^c}\right) + I\left(\mathbf{Y}_{\mathcal{S}};\hat{\mathbf{Y}}_{\mathcal{S}}|\mathbf{X}_{\mathcal{K}}\right) \nonumber\\
& \overset{(b)}{=} I\left(\mathbf{X}_{\mathcal{K}}; \hat{\mathbf{Y}}_{\mathcal{L}}\right) - I\left(\mathbf{X}_{\mathcal{K}};
\hat{\mathbf{Y}}_{\mathcal{S}^c}\right)+ \sum_{\ell \in \mathcal{S}}
I(\mathbf{Y}_{\ell};\hat{\mathbf{Y}}_{\ell}|\mathbf{X}_{\mathcal{K}})
\end{align}
for all $\mathcal{S}\subseteq\mathcal{L}$, where the equality (a) follows from the fact that
\begin{equation}\label{eqn:(a)-1}
I\left(\mathbf{Y}_{\mathcal{S}};\hat{\mathbf{Y}}_{\mathcal{S}}|\mathbf{X}_{\mathcal{K}},\hat{\mathbf{Y}}_{\mathcal{S}^c} \right) = I\left(\mathbf{Y}_{\mathcal{S}};\hat{\mathbf{Y}}_{\mathcal{S}}|\mathbf{X}_{\mathcal{K}}\right)
\end{equation}
and
\begin{equation}\label{eqn:(a)-2}
I\left(\mathbf{X}_{\mathcal{K}}; \hat{\mathbf{Y}}_{\mathcal{S}}| \mathbf{Y}_{\mathcal{S}}, \hat{\mathbf{Y}}_{\mathcal{S}^c}\right) = 0,
\end{equation}
and equality (b) follows from the fact that
\begin{equation}\label{eqn:(b)}
I\left(\mathbf{Y}_{\mathcal{S}};\hat{\mathbf{Y}}_{\mathcal{S}}|\mathbf{X}_{\mathcal{K}}\right) = \sum_{\ell \in \mathcal{S}}
I(\mathbf{Y}_{\ell};\hat{\mathbf{Y}}_{\ell}|\mathbf{X}_{\mathcal{K}}).
\end{equation}
The above equations (\ref{eqn:(a)-1})-(\ref{eqn:(b)}) follow from the Markov chain
\begin{equation*}
\hat{\mathbf{Y}}_{i} \leftrightarrow \mathbf{Y}_{i} \leftrightarrow
\mathbf{X}_{\mathcal{K}}
\leftrightarrow \mathbf{Y}_j \leftrightarrow \hat{\mathbf{Y}}_j, \quad \forall\; i\neq j.
\end{equation*}

We further evaluate the mutual information expression (\ref{eqn:multual_infor}) with Gaussian input and Gaussian quantization, which yields that
\begin{align*}
& I\left(\mathbf{Y}_{\mathcal{S}}; \hat{\mathbf{Y}}_{\mathcal{S}}| \hat{\mathbf{Y}}_{\mathcal{S}^c}\right) \\
& =  \log\frac{\left|\mathbf{H}_{\mathcal{L},\mathcal{K}} \mathbf{K}_{\mathcal{K}}
\mathbf{H}^{\dagger}_{\mathcal{L},\mathcal{K}}  + \mathrm{diag}\left(\{\mathbf{\Sigma}_{\ell} +
\mathbf{Q}_{\ell}\}_{\ell \in \mathcal{L}}\right) \right|}{\left|\mathrm{diag}\left(\{\mathbf{\Sigma}_{\ell} +
\mathbf{Q}_{\ell}\}_{\ell \in \mathcal{L}}\right)\right|} \\
& \quad - \log\frac{\left|\mathbf{H}_{\mathcal{S}^c,\mathcal{K}} \mathbf{K}_{\mathcal{K}}
\mathbf{H}^{\dagger}_{\mathcal{S}^c,\mathcal{K}}  + \mathrm{diag}\left(\{\mathbf{\Sigma}_{\ell} +
\mathbf{Q}_{\ell}\}_{\ell \in \mathcal{S}^c}\right) \right|}{\left|\mathrm{diag}\left(\{\mathbf{\Sigma}_{\ell} +
\mathbf{Q}_{\ell}\}_{\ell \in \mathcal{S}^c}\right)\right|}  \\
& \quad + \sum_{\ell\in \mathcal{S}}\log\frac{|\mathbf{\Sigma}_{\ell} +\mathbf{Q}_{\ell}|}{|\mathbf{Q}_{\ell}|}\\
& =  \log\frac{\left|\mathbf{H}_{\mathcal{L},\mathcal{K}} \mathbf{K}_{\mathcal{K}}
\mathbf{H}^{\dagger}_{\mathcal{L},\mathcal{K}}  + \mathrm{diag}\left(\{\mathbf{\Sigma}_{\ell} +
\mathbf{Q}_{\ell}\}_{\ell \in \mathcal{L}}\right) \right|}{\left|\mathbf{H}_{\mathcal{S}^c,\mathcal{K}} \mathbf{K}_{\mathcal{K}}
\mathbf{H}^{\dagger}_{\mathcal{S}^c,\mathcal{K}}  + \mathrm{diag}\left(\{\mathbf{\Sigma}_{\ell} +
\mathbf{Q}_{\ell}\}_{\ell \in \mathcal{S}^c}\right) \right|} \\
& \quad - \sum_{\ell\in \mathcal{S}}\log| \mathbf{Q}_{\ell}|  \\
& \leq \sum_{\ell\in \mathcal{S}}C_{\ell}.
\end{align*}

%

Instead of parameterizing the rate expressions over $\mathbf{Q}_{\ell}$ as in above, in this section, we introduce the following reparameterization, which is crucial for  proving our main results. Define
\begin{equation}
\label{eqn:reparameter-B}
\mathbf{B}_{\ell} = \left(\mathbf{\Sigma}_{\ell} + \mathbf{Q}_{\ell}\right)^{-1}.
\end{equation}
We represent the rate regions of joint decoding and successive decoding in terms of $\mathbf{B}_{\ell}$ in the following.

\begin{prop}
For the uplink C-RAN model shown in Fig.~\ref{fig:C-RAN-math} and under fixed Gaussian input $\mathbf{X}_\mathcal{K} \sim \mathcal{CN}(\mathbf{0}, \mathbf{K}_\mathcal{K})$ with $\mathbf{K}_\mathcal{K} = \mathrm{diag}\left(\{\mathbf{K}_k\}_{k \in \mathcal{K}}\right)$. The rate-fronthaul region for joint decoding under Gaussian quantization, $\mathcal{P}^G_{JD, GIn}$, is the closure of the convex hull of all $(R_1,\cdots,R_K, C_1,\ldots,C_L)$ satisfying
\begin{multline}\label{eqn:GaussRateRegion-JD}
\sum_{k\in \mathcal{T}} R_k <  \sum_{\ell\in \mathcal{S}}\left[C_{\ell} - \log \frac{|\mathbf{\Sigma}_{\ell}^{-1}|}{|\mathbf{\Sigma}_{\ell}^{-1} - \mathbf{B}_{\ell}|} \right] \\
 + \log\frac{\left|\sum_{\ell\in \mathcal{S}^c}
\mathbf{H}^{\dagger}_{\ell,\mathcal{T}} \mathbf{B}_{\ell} \mathbf{H}_{\ell,\mathcal{T}}  +
\mathbf{K}^{-1}_{\mathcal{T}}\right|}{\left|\mathbf{K}^{-1}_{\mathcal{T}}\right|}
\end{multline}
for all $\mathcal{T}\subseteq\mathcal{K}$ and $\mathcal{S}\subseteq\mathcal{L}$,
for some $0 \preceq \mathbf{B}_{\ell} \preceq \mathbf{\Sigma}^{-1}_{\ell}$, where
$\mathbf{K}_{\mathcal{T}} =
\mathbb{E}\left[\mathbf{X}_{\mathcal{T}}\mathbf{X}_{\mathcal{T}}^{\dagger}\right]$ is the
covariance matrix of $\mathbf{X}_{\mathcal{T}}$, and
$\mathbf{H}_{\ell,\mathcal{T}}$ denotes the channel matrix from
$\mathbf{X}_{\mathcal{T}}$ to $\mathbf{Y}_{\ell}$. Furthermore, under the fixed fronthaul capacity constraints $C_{\ell}$ for $\ell=1,\ldots,L$, the rate region achieved by joint decoding $\mathcal{R}^G_{JD,GIn}$ is defined as
\begin{align}\label{def:GaussRegion-JD}
\hspace{-1mm}\mathcal{R}^G_{JD,GIn} = \Big\{ & (R_1,\ldots,R_K): \nonumber\\
& (R_1,\cdots,R_K, C_1, \ldots, C_L) \in \mathcal{P}^G_{JD, GIn} \Big\}.
\end{align}
\end{prop}


\begin{prop}
For the uplink C-RAN model shown in Fig.~\ref{fig:C-RAN-math} and under fixed Gaussian input $\mathbf{X}_\mathcal{K} \sim \mathcal{CN}(\mathbf{0}, \mathbf{K}_\mathcal{K})$ with $\mathbf{K}_\mathcal{K} = \mathrm{diag}\left(\{\mathbf{K}_k\}_{k \in \mathcal{K}}\right)$. The rate-fronthaul region for successive decoding, $\mathcal{P}^G_{SD, GIn}$, is the closure of the convex hull of all $(R_1,\cdots,R_K, C_1,\ldots,C_L)$ satisfying
\begin{equation}
\label{eqn:GaussRateRegion-SD}
\displaystyle\sum_{k\in \mathcal{T}} R_k < 
\log\frac{\left|\sum_{\ell=1}^{L} \mathbf{H}^{\dagger}_{\ell,\mathcal{T}} \mathbf{B}_{\ell} \mathbf{H}_{\ell,\mathcal{T}}  + \mathbf{K}^{-1}_{\mathcal{T}}\right|}{\left|\mathbf{K}^{-1}_{\mathcal{T}}\right|},
\enspace \forall \; \mathcal{T}\subseteq\mathcal{K}, 
\end{equation}
and
\begin{multline}\label{eqn:GaussRegion-SD-2}
\displaystyle \log\frac{\left|\sum\limits_{\ell=1}^L \mathbf{H}^{\dagger}_{\ell,\mathcal{K}}
\mathbf{B}_{\ell} \mathbf{H}_{\ell,\mathcal{K}} +
\mathbf{K}_\mathcal{K}^{-1}\right|}{\left|\sum\limits_{\ell\in \mathcal{S}^c}
\mathbf{H}^{\dagger}_{\ell,\mathcal{K}} \mathbf{B}_{\ell} \mathbf{H}_{\ell, \mathcal{K}} + \mathbf{K}_\mathcal{K}^{-1} \right|} + \sum_{\ell\in \mathcal{S}}\log \frac{|\mathbf{\Sigma}_{\ell}^{-1}|}{|\mathbf{\Sigma}_{\ell}^{-1}-  \mathbf{B}_{\ell}|} \\
< \sum_{\ell\in \mathcal{S}}C_{\ell},
\enspace \forall \; \mathcal{S}\subseteq\mathcal{L},
\end{multline}
for some $0 \preceq \mathbf{B}_{\ell} \preceq \mathbf{\Sigma}^{-1}_{\ell}$, where
$\mathbf{K}_{\mathcal{T}} =
\mathbb{E}\left[\mathbf{X}_{\mathcal{T}}\mathbf{X}_{\mathcal{T}}^{\dagger}\right]$ is the
covariance matrix of $\mathbf{X}_{\mathcal{T}}$, and
$\mathbf{H}_{\ell,\mathcal{T}}$ denotes the channel matrix from
$\mathbf{X}_{\mathcal{T}}$ to $\mathbf{Y}_{\ell}$. Moreover, under the fixed fronthaul capacity constraints $C_{\ell}$ for $\ell=1,\ldots,L$, the rate region achieved by successive decoding $\mathcal{R}^G_{SD,GIn}$ is defined as
\begin{align}\label{def:GaussRegion-SD}
\hspace{-1mm}\mathcal{R}^G_{SD,GIn} = \Big\{ & (R_1,\ldots,R_K): \nonumber \\
&(R_1,\cdots,R_K, C_1, \ldots, C_L) \in \mathcal{P}^G_{SD, GIn}\Big\}.
\end{align}
\end{prop}

\subsection{Gaussian Input and Gaussian Quantization Achieve Capacity to within Constant Gap}
\label{subsec:ConstantGap-Gaussian}

With Gaussian input and Gaussian quantization, the rate region of joint decoding (\ref{eqn:GaussRateRegion-JD}) can be shown to be within a constant gap to the capacity region of uplink C-RAN. This constant-gap result is stated in the following theorem.

\begin{thm}
\label{thm:constant-gap-JD}
For any rate tuple $(R_1, R_2, \ldots, R_K)$ within the cut-set bound for uplink
C-RAN with fixed fronthaul capacities of $C_{\ell}$ shown in Fig.~\ref{fig:C-RAN-math}, the rate tuple $(R_1 -  \eta, R_2- \eta, \ldots, R_K- \eta)$, with $\eta = NL + M$ is achievable for compress-and-forward with Gaussian input, Gaussian quantization, and joint decoding, where $L$ is the number of BSs in the network, $M$ is the number of transmit antennas at user, and $N$ is the number of receive antennas at BS, i.e., $(R_1 -  \eta, R_2- \eta, \ldots, R_K- \eta) \in \mathcal{R}^G_{JD,GIn}$.
\end{thm}

\begin{IEEEproof}
See Appendix~\ref{append:proof-JD-cons-gap}.
\end{IEEEproof}

Although the uplink C-RAN model is an example of a relay network for which noisy network coding approach applies and
it is known that compress-and-forward with joint decoding achieves the same rate region as noisy network coding for uplink C-RAN, we remark that Theorem~\ref{thm:constant-gap-JD} does not immediately follow from the constant-gap optimality result of noisy network coding~\cite{Lim11}. The constant-gap optimality of noisy network coding is proven for Gaussian relay networks, whereas the uplink C-RAN model contains fronthaul links which are digital connections and not Gaussian channels.

Combining with our earlier results on the optimality of successive decoding, constant-gap optimality results can also be obtained for compress-and-forward with generalized successive decoding and successive decoding. These results are summarized in the following corollary.


\begin{cor}\label{cor:SD-ConstantGap}
For the uplink C-RAN model as shown in Fig.~\ref{fig:C-RAN-math}, compress-and-forward with generalized successive decoding, under Gaussian input and Gaussian quantization achieves the capacity region to within $NL + M$ bits per complex dimension if the fronthaul links are subjected to a sum capacity constraint $\sum_{\ell=1}^L C_{\ell} \leq C$. Furthermore, compress-and-forward with successive decoding, under Gaussian input and Gaussian quantization, achieves the sum capacity of an uplink C-RAN model with individual fronthaul constraints to within $NL + MK$ bits per complex dimension.
\end{cor}


\subsection{Optimality of Gaussian Quantization under Joint Decoding}
\label{subsec:opt-Gaussian}

For the Gaussian uplink MIMO C-RAN model, it is known that
Gaussian input and Gaussian quantization are not jointly optimal~\cite{Sand08}. However, if the quantization noise is
fixed as Gaussian, then the optimal input distribution must be Gaussian.
This is because the channel reduces to a conventional Gaussian
multiple-access channel in this case. The main result of this section is that the converse is also true, i.e., under fixed Gaussian input, Gaussian quantization actually maximizes the achievable rate region of the uplink C-RAN model under joint decoding.

Under fixed fronthaul capacity constraints $C_{\ell}$ for $\ell=1,\ldots,L$, we let $\mathcal{R}^*_{JD, GIn}$ denote the rate region of joint decoding under Gaussian input and optimal quantization.
In the following, we first define Fisher information and state the two main tools for proving this result: the Bruijn identity and the Fisher information inequality. We then present the main theorem on the optimality of Gaussian quantization for joint decoding, i.e., $\mathcal{R}^G_{JD, GIn} = \mathcal{R}^*_{JD, GIn}$.

\begin{defn}
Let $\left(\mathbf{X}, \mathbf{Y}\right)$ be a pair of random vectors with joint probability distribution function $p\left(\mathbf{x}, \mathbf{y}\right)$. The Fisher information matrix of $\mathbf{X}$ is defined as
\begin{equation}
\mathbf{J}\left(\mathbf{X}\right) = \mathbb{E}\left[\nabla \log p\left(\mathbf{X}\right) \nabla \log p\left(\mathbf{X}\right)^{\mathsf{T}}\right].
\end{equation}
Likewise, the Fisher information matrix of $\mathbf{X}$ conditional on $\mathbf{Y}$ is defined as
\begin{equation}
\mathbf{J}\left(\mathbf{X}| \mathbf{Y}\right) = \mathbb{E}\left[\nabla \log p\left(\mathbf{X}|\mathbf{Y}\right)\nabla \log p\left(\mathbf{X}|\mathbf{Y}\right)^{\mathsf{T}}  \right].
\end{equation}
\end{defn}

\begin{lemma}[{Fisher Information Inequality, \cite{Dembo91}~\cite[Lemma 2]{EU14}}]
\label{lem:Fisher-entropy}
Let $(\mathbf{U},\mathbf{X})$ be an arbitrary complex random vector, where the conditional Fisher information of $\mathbf{X}$ conditioned on $\mathbf{U}$ exists. We have
\begin{equation}
\log \left|(\pi e) \mathbf{J}^{-1}\left(\mathbf{X}|\mathbf{U}\right)\right| \leq h\left(\mathbf{X}|\mathbf{U}\right).
\end{equation}
\end{lemma}

\begin{lemma}[{Bruijn Identity, \cite{Palomar06}~\cite[Lemma 3]{EU14}}]\label{lem:Fisher-MMSE}
Let $(\mathbf{V}_1,\mathbf{V}_2)$ be an arbitrary random vector with finite second moments, and $\mathbf{N}$ be a zero-mean Gaussian random vector with covariance $\mathbf{\Lambda}_N$. Assume
$(\mathbf{V}_1,\mathbf{V}_2)$ and $\mathbf{N}$ are independent. We have
\begin{equation}
\cov\left(\mathbf{V}_2 | \mathbf{V}_1, \mathbf{V}_2 + \mathbf{N} \right) = \mathbf{\Lambda}_N - \mathbf{\Lambda}_N \mathbf{J}\left(\mathbf{V}_2 + \mathbf{N} | \mathbf{V}_1\right)\mathbf{\Lambda}_N.
\end{equation}
\end{lemma}

\begin{thm}\label{thm:GaussianOpt-JD}
For the uplink C-RAN under fixed Gaussian input distribution and assuming joint
decoding, Gaussian quantization is optimal, i.e., $\mathcal{R}^G_{JD, GIn} = \mathcal{R}^*_{JD, GIn}$.
\end{thm}

\begin{IEEEproof}
Recall that the achievable rate region of the compress-and-forward scheme under joint decoding is given by the set of $(R_1,\ldots,R_K)$ derived from (\ref{eqn:Rate-JD})
under the joint distribution
\begin{multline}
p\left(\mathbf{x}_1,\ldots, \mathbf{x}_K, \mathbf{y}_1, \ldots, \mathbf{y}_L, \hat{\mathbf{y}}_1, \ldots, \hat{\mathbf{y}}_L\right) \\
= \prod_{k=1}^Kp\left(\mathbf{x}_k\right)\prod_{\ell=1}^Lp\left(\mathbf{y}_{\ell}|\mathbf{x}_1,\ldots, \mathbf{x}_K\right) \prod_{\ell=1}^Lp\left(\hat{\mathbf{y}}_{\ell}|\mathbf{y}_{\ell}\right).
\end{multline}
For fixed Gaussian input $\mathbf{X}_\mathcal{K} \sim \mathcal{CN}(\mathbf{0}, \mathbf{K}_\mathcal{K})$ and fixed $\prod_{\ell=1}^Lp(\hat{\mathbf{y}}_{\ell}|\mathbf{y}_{\ell})$, choose
$\mathbf{B}_{\ell}$ with $\mathbf{0} \preceq \mathbf{B}_{\ell} \preceq \mathbf{\Sigma}_{\ell}^{-1}$ such that
\begin{align*}
 \cov\left(\mathbf{Y}_{\ell} | \mathbf{X}_{\mathcal{K}}, \mathbf{\hat{Y}}_{\ell}\right) = \mathbf{\Sigma}_{\ell} - \mathbf{\Sigma}_{\ell} \mathbf{B}_{\ell}\mathbf{\Sigma}_{\ell},\quad \ell=1,\cdots,L.
\end{align*}
We proceed to show that the achievable rate region as given by
(\ref{eqn:GaussRateRegion-JD}) with a Gaussian
$\prod_{\ell=1}^Lp(\hat{\mathbf{y}}_{\ell}|\mathbf{y}_{\ell}) \sim \mathcal{CN}(\mathbf{Y}_{\ell}, \mathbf{Q}_{\ell})$,
where $\mathbf{Q}_{\ell} = \mathbf{B}_{\ell}^{-1} - \mathbf{\Sigma}_{\ell}$,
is as large as that of (\ref{eqn:Rate-JD}) under Gaussian input.

First, note that
\begin{align}
I\left(\mathbf{Y}_{\ell}; \hat{\mathbf{Y}}_{\ell} | \mathbf{X}_{\mathcal{K}}\right)& = \log\left|(\pi
e)\mathbf{\Sigma}_{\ell}\right| - h\left(\mathbf{Y}_{\ell} | \mathbf{X}_{\mathcal{K}},
\hat{\mathbf{Y}}_{\ell}\right) \nonumber \\
& \geq\log\left|(\pi e)\mathbf{\Sigma}_{\ell}\right|-\log\left|(\pi e)\cov\left(\mathbf{Y}_{\ell} | \mathbf{X}_{\mathcal{K}}, \mathbf{\hat{Y}}_{\ell}\right)\right| \nonumber \\
&=\log \frac{\left|\mathbf{\Sigma}_{\ell}^{-1}\right|}{\left|\mathbf{\Sigma}_{\ell}^{-1}-  \mathbf{B}_{\ell}\right|},\quad \ell=1,\cdots,L,
\label{eqn:JC_Proof_first_term}
\end{align}
where we use the fact that Gaussian distribution maximizes differential
entropy.

Moreover, we have
\begin{align*}
I\left(\mathbf{X}_{\mathcal{T}};\hat{\mathbf{Y}}_{\mathcal{S}^c}|\mathbf{X}_{\mathcal{T}^c}\right) &= h\left(\mathbf{X}_{\mathcal{T}}\right) - h\left(\mathbf{X}_{\mathcal{T}}| \mathbf{X}_{\mathcal{T}^{c}},\hat{\mathbf{Y}}_{\mathcal{S}^c}\right)\\
&\leq  \log \left|\mathbf{K}_{\mathcal{T}}\right| -  \log\left| \mathbf{J}^{-1}\left(\mathbf{X}_{\mathcal{T}}|\mathbf{X}_{\mathcal{T}^c},\hat{\mathbf{Y}}_{\mathcal{S}^c}\right)\right|,
\end{align*}
where the inequality is due to Lemma~\ref{lem:Fisher-entropy}. Since
\begin{align*}
\mathbf{Y}_{\mathcal{S}^c} & = \mathbf{H}_{\mathcal{S}^c, \mathcal{T}} \mathbf{X}_{\mathcal{T}} + \mathbf{H}_{\mathcal{S}^c, \mathcal{T}^{c}} \mathbf{X}_{\mathcal{T}^{c}}+\mathbf{Z}_{\mathcal{S}^c} ,
\end{align*}
it follows from the MMSE estimation of Gaussian random vectors that
\begin{align*}
\mathbf{X}_{\mathcal{T}}&=\mathbb{E}\left[\mathbf{X}_{\mathcal{T}}|\mathbf{X}_{\mathcal{T}^c},\mathbf{Y}_{\mathcal{S}^c}\right]+\mathbf{N}_{\mathcal{T},\mathcal{S}^c}\\
&=\sum_{\ell \in \mathcal{S}^c}\mathbf{G}_{\mathcal{T},\ell} \left(\mathbf{Y}_{\ell}-\mathbf{H}_{\ell, \mathcal{T}^{c}} \mathbf{X}_{\mathcal{T}^{c}}\right)+\mathbf{N}_{\mathcal{T},\mathcal{S}^c},
\end{align*}
where 
\begin{equation*}
\mathbf{G}_{\mathcal{T}, \ell} = \left(\mathbf{K}_{\mathcal{T}}^{-1} + \sum_{j \in \mathcal{S}^c}  \mathbf{H}^{\dagger} _{j, \mathcal{T} }\mathbf{\Sigma}_{j}^{-1}\mathbf{H}_{j,\mathcal{T}}\right)^{-1}\mathbf{H}^{\dagger}_{\ell, \mathcal{T}} \mathbf{\Sigma}_{\ell}^{-1},
\end{equation*}
and $\mathbf{N}_{\mathcal{T},\mathcal{S}^c} \sim \mathcal{CN} \left(\mathbf{0},  \mathbf{\Lambda}_{\mathbf{N}}  \right)$
with covariance matrix
\begin{equation}
\mathbf{\Lambda}_{\mathbf{N}} = \left(\mathbf{K}_{\mathcal{T}}^{-1} + \sum_{\ell \in \mathcal{S}^c}  \mathbf{H}^{\dagger} _{\ell, \mathcal{T} }\mathbf{\Sigma}_{\ell}^{-1}\mathbf{H}_{\ell,\mathcal{T}} \right)^{-1}.
\end{equation}
Here $\mathbb{E}\left[\mathbf{X}_{\mathcal{T}}|\mathbf{X}_{\mathcal{T}^c},\mathbf{Y}_{\mathcal{S}^c}\right]$ is the MMSE estimator of $\mathbf{X}_{\mathcal{T}}$ from $\mathbf{X}_{\mathcal{T}^c},\mathbf{Y}_{\mathcal{S}^c}$. The error in estimation is $\mathbf{N}_{\mathcal{T},\mathcal{S}^c}$, and the MMSE matrix is $\mathbf{\Lambda}_{\mathbf{N}}$.

By the matrix complementary identity between Fisher information matrix and MMSE in Lemma~\ref{lem:Fisher-MMSE}, we have
\begin{align*}
& \mathbf{J}\left(\mathbf{X}_{\mathcal{T}}|\mathbf{X}_{\mathcal{T}^c},\hat{\mathbf{Y}}_{\mathcal{S}^c}\right) \\
& = \mathbf{\Lambda}_{\mathbf{N}}^{-1} \\
& \enspace - \mathbf{\Lambda}_{\mathbf{N}}^{-1}
\cov \left( \sum_{\ell \in \mathcal{S}^c}\mathbf{G}_{\mathcal{T},\ell} (\mathbf{Y}_{\ell}-\mathbf{H}_{\ell,\mathcal{T}^{c}} \mathbf{X}_{\mathcal{T}^{c}}) | \mathbf{X}_{\mathcal{K}},
\hat{\mathbf{Y}}_{\mathcal{S}^c} \right) \mathbf{\Lambda}_{\mathbf{N}}^{-1}   \\
& =  \mathbf{\Lambda}_{\mathbf{N}}^{-1}   - \mathbf{\Lambda}_{\mathbf{N}}^{-1} \cov \left(
\sum_{\ell \in \mathcal{S}^c}\mathbf{G}_{\mathcal{T},\ell} \mathbf{Y}_{\ell} |
\mathbf{X}_{\mathcal{K}}, \hat{\mathbf{Y}}_{\mathcal{S}^c}\right)\mathbf{\Lambda}_{\mathbf{N}}^{-1}   \\
& =  \mathbf{\Lambda}_{\mathbf{N}}^{-1}  - \mathbf{\Lambda}_{\mathbf{N}}^{-1}\left[\sum_{\ell \in \mathcal{S}^c}\mathbf{G}_{\mathcal{T},\ell}\cov \left( \mathbf{Y}_{\ell} | \mathbf{X}_{\mathcal{K}}, \hat{\mathbf{Y}}_{\ell}\right)\mathbf{G}^{\dagger}_{\mathcal{T},\ell}\right]\mathbf{\Lambda}_{\mathbf{N}}^{-1}   \\
& = \mathbf{\Lambda}_{\mathbf{N}}^{-1}  - \sum_{\ell \in \mathcal{S}^c}\mathbf{H}^{\dagger}_{\ell, \mathcal{T}}        \left(\mathbf{\Sigma}_{\ell}^{-1}-\mathbf{B}_{\ell} \right)       \mathbf{H}_{\ell, \mathcal{T}} \\
& =\mathbf{K}_{\mathcal{T}}^{-1} + \sum_{\ell \in \mathcal{S}^c} \mathbf{H}^{\dagger}_{\ell, \mathcal{T}} \mathbf{B}_{\ell}\mathbf{H}_{\ell, \mathcal{T}}.
\end{align*}

Therefore,
\begin{multline}\label{eqn:JC_Proof_second_term}
I\left(\mathbf{X}_{\mathcal{T}};\hat{\mathbf{Y}}_{\mathcal{S}^c}|\mathbf{X}_{\mathcal{T}^c}\right)  \leq
\log\frac{\left| \mathbf{J}(\mathbf{X}_{\mathcal{T}}|\mathbf{X}_{\mathcal{T}^c},\hat{\mathbf{Y}}_{\mathcal{S}^c})\right|}{\left| \mathbf{K}^{-1}_{\mathcal{T}}\right|}  \\
 = \log\frac{\left|\mathbf{K}_{\mathcal{T}}^{-1} + \sum_{\ell \in \mathcal{S}^c} \mathbf{H}^{\dagger}_{\ell, \mathcal{T}} \mathbf{B}_{\ell}\mathbf{H}_{\ell, \mathcal{T}}\right|}{\left|\mathbf{K}_{\mathcal{T}}^{-1}\right|}
\end{multline}
for all  $\mathcal{T}\subseteq\mathcal{K}$ and $\mathcal{S}\subseteq\mathcal{L}$.
Combining (\ref{eqn:JC_Proof_first_term}) and (\ref{eqn:JC_Proof_second_term}), we
conclude that $\mathcal{R}_{JD,GIn}^G$ as derived from (\ref{eqn:GaussRateRegion-JD}) is as large as
$\mathcal{R}_{JD, GIn}^*$. Therefore, $\mathcal{R}_{JD, GIn}^G = \mathcal{R}_{JD, GIn}^*$.
\end{IEEEproof}


\subsection{Optimization of Gaussian Input and Gaussian Quantization Noise Covariance Matrices}
\label{chapOpt-sec-ConvOpt}

This section addresses the numerical optimization of the Gaussian input and quantization noise covariance matrices for uplink MIMO C-RAN under given fronthaul capacity constraints. First, we note that even when restricting to Gaussian input and Gaussian quantization,
the joint optimization of input and quantization noise covariance
matrices is still a challenging problem for the uplink MIMO C-RAN.
However, if we fix the quantization noise covariance, then the
input optimization reduces to that of optimizing a conventional
Gaussian multiple-access channel. In particular, the problem of
maximizing the weighted sum rate can be formulated as a convex
optimization, which can be readily solved \cite{yu2004iterative}.

Conversely, if we fix the transmit covariance matrix, the optimization
of quantization noise covariance can in some cases be formulated as convex
optimization. The key enabling fact is the reparameterization in term
of $\mathbf{B}_{\ell}$ (\ref{eqn:reparameter-B}), instead of direct optimization over
$\mathbf{Q}_{\ell}$.  Consider first the case of joint decoding.
Using (\ref{eqn:GaussRateRegion-JD}) under the fixed $C_{\ell}$ for $\ell=1,\ldots,L$,
the weighted sum rate maximization problem can be formulated over
$\{R_k, \mathbf{B}_{\ell}\}$ as follows:
\begin{eqnarray}\label{eqn:JDWeighSum-COV}
\hspace{-3mm}& \displaystyle \max_{R_k, \mathbf{B}_{\ell}} &  \sum_{k=1}^K \mu_k R_k \\
\hspace{-3mm}& \mathrm{s.t.} & \sum_{k\in \mathcal{T}} R_k \leq  \log\frac{\left|\sum_{\ell\in \mathcal{S}^c}
\mathbf{H}^{\dagger}_{\ell,\mathcal{T}} \mathbf{B}_{\ell} \mathbf{H}_{\ell,\mathcal{T}}  +
\mathbf{K}^{-1}_{\mathcal{T}}\right|}{\left|\mathbf{K}^{-1}_{\mathcal{T}}\right|}   \nonumber \\
\hspace{-3mm}& & \, + \sum_{\ell\in \mathcal{S}}\left[C_{\ell} - \log \frac{|\mathbf{\Sigma}_{\ell}^{-1}|}{|\mathbf{\Sigma}_{\ell}^{-1} - \mathbf{B}_{\ell}|} \right], \;\forall \; \mathcal{T}\subseteq \mathcal{K}, \; \forall \mathcal{S}\subseteq \mathcal{L}, \nonumber \\
\hspace{-3mm}& & \mathbf{0} \preceq \mathbf{B}_{\ell} \preceq \mathbf{\Sigma}_{\ell}^{-1}, \quad \forall \; \ell\in\mathcal{L}, \nonumber
\end{eqnarray}
where $\mu_k$ represents the weight associated with user $k$, which is
typically determined from upper layer protocols. The key observation
is that the above problem is convex in $\{R_k, \mathbf{B}_{\ell}\}$.
However, we also note that because of joint decoding, the number of
constraints is exponential in the size of the network. Consequently,
the above optimization problem can only be solved for small networks
in practice.

Note that the above formulation considers the optimization of instantaneous achievable rates $R_k$ under instantaneous fronthaul capacity constraints $C_{\ell}$ in a fixed time slot. The solution obtained, however, also applies to the more general case of optimizing the weighted sum rates under weighted sum fronthaul constraint (e.g., $\sum_{\ell=1}^{L} \nu_{\ell} C_{\ell} \leq C$). This is because if we consider a slightly more general formulation of optimizing an objective of
\begin{equation}\label{eqn:weighted-sum-fronthaul}
\max_{R_k,\mathbf{B}_{\ell}, C_{\ell}} \sum_{k=1}^K \mu_k R_k - \gamma \sum_{\ell=1}^L \nu_{\ell} C_{\ell}
\end{equation}
under the same constraints as in (\ref{eqn:JDWeighSum-COV}) and $\sum_{\ell=1}^{L} \nu_{\ell} C_{\ell} \leq C$. Such an optimization problem is convex, so time-sharing is not needed. For this reason, the rest of this section considers the formulation with instantaneous rates only.

We now consider the weighted sum-rate maximization problem for the
case of successive decoding of the quantization codewords followed by
the user messages. However, the direct characterization of successive decoding
rate (\ref{eqn:GaussRateRegion-SD})-(\ref{eqn:GaussRegion-SD-2}) does not give rise to a convex formulation.
Nevertheless, for the special case of maximizing the sum rate (i.e., with
$\mu_1 = \cdots = \mu_K=1$), using Theorem~\ref{thm:opt-VMAC}, which shows
that successive decoding achieves the same maximum sum rate as joint decoding,
the sum-rate maximization problem with successive decoding can be equivalently
formulated as follows:

\begin{thm}\label{thm:SumRate-COV}
For the uplink C-RAN model with individual fronthaul capacity constraint $C_{\ell}$ as shown in Fig.~\ref{fig:C-RAN-math}, the sum rate maximization problem under successive decoding can be formulated as the following convex problem:
\begin{eqnarray}\label{eqn:Sum-SD-COV}
& \displaystyle \max_{R, \mathbf{B}_{\ell}} &   R \\
& \mathrm{s.t.} &  R \leq  \sum_{\ell\in \mathcal{S}}\left[C_{\ell} - \log \frac{|\mathbf{\Sigma}_{\ell}^{-1}|}{|\mathbf{\Sigma}_{\ell}^{-1} - \mathbf{B}_{\ell}|} \right]  \nonumber \\
&& \qquad + \log\frac{\left|\sum_{\ell\in \mathcal{S}^c}
\mathbf{H}^{\dagger}_{\ell,\mathcal{T}} \mathbf{B}_{\ell} \mathbf{H}_{\ell,\mathcal{T}}  +
\mathbf{K}^{-1}_{\mathcal{K}}\right|}{\left|\mathbf{K}^{-1}_{\mathcal{K}}\right|},  \enspace \forall \mathcal{S}\subseteq \mathcal{L}, \nonumber \\
&& \mathbf{0} \preceq \mathbf{B}_{\ell} \preceq \mathbf{\Sigma}_{\ell}^{-1}, \enspace \forall \; \ell\in \mathcal{L}. \nonumber
\end{eqnarray}
Further, if the fronthaul links are subject to a sum capacity
constraint of $C$, the sum rate maximization problem can be formulated as the following convex problem:
\begin{eqnarray}\label{eqn:SumFront-SD-COV}
& \displaystyle \max_{R, \mathbf{B}_{\ell}} &   R \\
& \mathrm{s.t.} &  R \leq 
\log\frac{\left|\sum_{\ell=1}^{L} \mathbf{H}^{\dagger}_{\ell,\mathcal{K}} \mathbf{B}_{\ell} \mathbf{H}_{\ell,\mathcal{K}}  + \mathbf{K}^{-1}_{\mathcal{K}}\right|}{\left|\mathbf{K}^{-1}_{\mathcal{K}}\right|}, \nonumber \\
&&  R  + \sum_{\ell=1}^L\log
\frac{|\mathbf{\Sigma}_{\ell}^{-1}|}{|\mathbf{\Sigma}_{\ell}^{-1}-
\mathbf{B}_{\ell}|} \leq C, \nonumber \\
& & \mathbf{0} \preceq \mathbf{B}_{\ell} \preceq \mathbf{\Sigma}_{\ell}^{-1}, \enspace \forall\; \ell\in \mathcal{L}. \nonumber
\end{eqnarray}
\end{thm}

We remark that the formulation for
uplink C-RAN with individual fronthaul capacities (\ref{eqn:Sum-SD-COV})
has exponential number of constraints, because the CP in effect needs
to search over $L!$ different decoding orders of quantization
codewords at the BSs. In practical implementation, a heuristic method
can be used to determine the decoding orders of quantization
codewords for avoiding the exponential search~\cite{park2014fronthaul,zhou2015SP}.
Alternatively, if the C-RAN has a sum fronthaul
constraint, then the number of constraints is linear in network size,
because we only need to consider the case of $\mathcal{S} = \mathcal{L}$
and $\mathcal{S}=\emptyset$ in (\ref{eqn:Sum-SD-COV}).
Consequently, the resulting quantization noise covariance optimization problem
(\ref{eqn:SumFront-SD-COV}) can be solved in polynomial time. Note that convexity is a key advantage of the above problem formulations as
compared to previous approaches in the literature (e.g. \cite{Del09,
Park12}) that parameterize the optimization problem over the
quantization noise covariance $\mathbf{Q}_{\ell}$, which leads to a
nonconvex formulation.

We emphasize the importance of Gaussian input for the convex formulation in Theorem~\ref{thm:SumRate-COV}. Suppose that both input signal $\mathbf{X}_{\mathcal{K}}$ and compressed signal $\hat{\mathbf{Y}}_{\ell}$ are discrete random vectors with finite alphabet. For fixed input distribution, the sum-rate maximization problem under the sum fronthaul constraint can be written as
\begin{eqnarray}\label{eqn:RateMax-nonGauss}
& \displaystyle \max_{p(\hat{\mathbf{y}}_{\ell}|\mathbf{y}_{\ell})}& I\left(\mathbf{X}_{\mathcal{K}}; \hat{\mathbf{Y}}_{\mathcal{L}}\right), \\ 
& \mathrm{s.t.} & I\left(\mathbf{Y}_{\mathcal{L}}; \hat{\mathbf{Y}}_{\mathcal{L}}\right) \leq C, \nonumber \\
& & p\left(\hat{\mathbf{y}}_{\ell}|\mathbf{y}_{\ell}\right) \geq 0, \enspace \sum_{\hat{\mathbf{y}}_{\ell}} p\left(\hat{\mathbf{y}}_{\ell}|\mathbf{y}_{\ell}\right) =1,  \enspace \forall\; \ell\in \mathcal{L}. \nonumber
\end{eqnarray}
The above problem can be thought as a variant of the information bottleneck method~\cite{tishby1999information}, which can be solved by a generalized Blahut-Arimoto (BA) algorithm~\cite{blahut1972computation, arimoto1973converse}. However, due to the non-convex nature of problem (\ref{eqn:RateMax-nonGauss}), the generalized BA algorithm can only converge to a local optimum.

\section{Conclusion}
\label{chapOpt-sec-conclustion}

This paper provides a number of information theoretical results on the optimal
compress-and-forward scheme for the uplink MIMO C-RAN model, where the
BSs are connected to a CP through noiseless fronthaul links of limited
capacities. It is shown that the generalized successive decoding
scheme, which allows arbitrary decoding orders between quantization
and message codewords, can achieve the same rate region as joint
decoding under a sum fronthaul constraint.  Moreover, the practical
successive decoding of the quantization codewords followed by the user
messages is shown to achieve the same maximum sum rate as joint
decoding under individual fronthaul constraints.  In addition, if the
input distribution is assumed to be Gaussian, it is shown that Gaussian
quantization maximizes the achievable rate region of joint decoding.
With Gaussian input signaling, the optimization of Gaussian
quantization for maximizing the weighted sum rate under joint decoding
and the sum rate under successive decoding can be cast as convex
optimization problems, which facilitates efficient numerical
solution. Finally, Gaussian input and Gaussian quantization achieve
the capacity region of the uplink C-RAN model to within constant gap.
Collectively, these results provide justifications for the practical
choice of using Gaussian-like input signals at the user terminals, Gaussian-like
quantization at the relaying BSs, and successive decoding of
quantization codewords followed by user messages at the CP for
implementing uplink MIMO C-RAN.

\appendices

\section{Optimality of Generalized Successive Decoding}
\label{append:proof-GSD=JD}

In this appendix, we prove Theorem~\ref{thm:GSD=JD-SumFront}, which states the equivalence between generalized successive decoding and joint decoding under a sum-capacity fronthaul constraint. We begin by introducing an outer bound for the achievable rate region of joint decoding under a sum fronthaul constraint. Under the sum fronthaul capacity constraint,
define the rate-fronthaul region for joint decoding $\mathcal{P}^{o}_{JD,s}$ as the closure of the convex hull of all $(R_1, R_2, \ldots, R_K, C)$ satisfying
\begin{equation}\label{eqn:RateRegion-JD-SumFront}
\begin{cases}
\enspace \displaystyle \sum_{k\in \mathcal{T}}R_{k} < \min\bigg\{ C - \sum_{\ell\in \mathcal{L}}I\left(\mathbf{Y}_{\ell};\hat{\mathbf{Y}}_{\ell}|\mathbf{X}_{\mathcal{K}}\right),  \\
\qquad \qquad \qquad \qquad \quad \; I\left(\mathbf{X}_{\mathcal{T}}; \hat{\mathbf{Y}}_{\mathcal{L}}|\mathbf{X}_{\mathcal{T}^c}\right)\bigg\}, \quad \forall \; \mathcal{T}\subseteq \mathcal{K}, \\
\enspace C > \sum_{\ell\in \mathcal{L}}I\left(\mathbf{Y}_{\ell};\hat{\mathbf{Y}}_{\ell}|\mathbf{X}_{\mathcal{K}}\right)
\end{cases}
\end{equation}
for some product distribution $\prod_{k=1}^Kp\left(\mathbf{x}_k\right)\prod_{\ell=1}^Lp(\hat{\mathbf{y}}_{\ell}|\mathbf{y}_{\ell})$. Under fixed sum fronthaul constraint $C$, define the region $\mathcal{R}^{o}_{JD,s}$ as follows
\begin{equation}
\label{def:JD-outerbound}
\mathcal{R}^{o}_{JD,s} = \Big\{(R_1,\ldots,R_K): (R_1,\cdots,R_K, C) \in \mathcal{P}^{o}_{JD,s}\Big\}.
\end{equation}
Note that the rate region $\mathcal{R}^o_{JD,s}$ is an outer bound for joint decoding rate region (\ref{def:Region-JD-SumFront}) because only the constraints corresponding to $\mathcal{S}=\emptyset$ and $\mathcal{S} = \mathcal{L}$ are included. These constraints turn out to be the only active ones under the sum fronthaul constraint $\sum_{\ell=1}^LC_{\ell} \leq C$ and $C_{\ell}\geq 0$.
Under the sum fronthaul constraint, the generalized successive decoding region $\mathcal{P}_{GSD,s}(\pi)$ for decoding order $\pi$ can be derived from (\ref{eqn:Rate-JD}) by letting $\sum_{\ell=1}^LC_{\ell} = C$. More specifically, $\mathcal{P}_{GSD,s}(\pi)$ is the closure of the convex hull of all $(R_1, R_2, \ldots, R_K, C)$ satisfying 
\begin{equation}\label{eqn:RateRegion-GSD-SumFront}
\begin{cases}
\enspace \displaystyle R_{k} <I\left( \mathbf{X}_{k}; \hat{\mathbf{Y}}_{\mathcal{J}_{\mathbf{X}_k}}|\mathbf{X}_{\mathcal{I}_{\mathbf{X}_k}}\right),  \quad \forall \; k\in \mathcal{K}, \\
\enspace \displaystyle C > \sum_{\ell=1}^L I\left( \mathbf{Y}_{\ell}; \hat{\mathbf{Y}}_{\ell}|\hat{\mathbf{Y}}_{\mathcal{J}_{\mathbf{Y}_{\ell}}},\mathbf{X}_{\mathcal{I}_{\mathbf{Y}_{\ell}}}\right), 
\end{cases}
\end{equation}
for some product distribution $\prod_{k=1}^Kp\left(\mathbf{x}_k\right)\prod_{\ell=1}^Lp(\hat{\mathbf{y}}_{\ell}|\mathbf{y}_{\ell})$, where $\mathcal{I}_{\mathbf{X}_k}$, $\mathcal{I}_{\mathbf{Y}_{\ell}}$ are the indices of user messages that are decoded before $\mathbf{X}_k$ and $\mathbf{Y}_{\ell}$ under the permutation $\pi$, and $\mathcal{J}_{\mathbf{X}_k}$, $\mathcal{J}_{\mathbf{Y}_{\ell}}$ are the indices of the quantization codewords that are decoded before $\mathbf{X}_k$ and $\mathbf{Y}_{\ell}$ under decoding order $\pi$. Define $\mathcal{P}^*_{GSD,s}$ to be the closure of the convex hull of all $\mathcal{P}_{GSD,s}(\pi)$'s over decoding order $\pi$'s, i.e., $\mathcal{P}^*_{GSD,s}= \mathrm{co}\left( \bigcup\limits_{\pi}\mathcal{P}_{GSD,s}(\pi)\right)$.

We say a point $(R_1, \ldots, R_K, C)$ is \emph{dominated} by a point in $\mathcal{P}^*_{GSD,S}$ if there exists some $(R'_1, \ldots, R'_K, C')$ in $\mathcal{P}^*_{GSD,s}$ for which $R_k \leq R'_k$ for $k=1,2,\ldots,K$, and $C\geq C'$.

Given the definitions of $\mathcal{R}^*_{GSD,s}$, $\mathcal{R}^*_{JD,s}$ and $\mathcal{R}^o_{JD,s}$, it is easy to see that $\mathcal{R}^*_{GSD,s}\subseteq \mathcal{R}^*_{JD,s}\subseteq \mathcal{R}^o_{JD,s}$.  To show $\mathcal{R}^*_{GSD,s}= \mathcal{R}^*_{JD,s}$, it suffices to show  $\mathcal{R}^o_{JD,s}\subseteq \mathcal{R}^*_{GSD,s}$, which is equivalent to showing that if a point $(R_1, R_2, \ldots, R_K, C)\in \mathcal{P}^o_{JD,s}$, then the same point $(R_1, R_2, \ldots, R_K, C) \in \mathcal{P}^*_{GSD,s}$ also. To show this, it suffices to show that for any fixed product distribution $\prod_{k=1}^Kp\left(\mathbf{x}_k\right)\prod_{\ell=1}^Lp(\hat{\mathbf{y}}_{\ell}|\mathbf{y}_{\ell})$ and fixed $C$, each extreme point $(R_1, \ldots, R_K, C)$ as defined by (\ref{eqn:RateRegion-JD-SumFront}) is dominated by a point in $\mathcal{P}^*_{GSD,s}$ with the average sum fronthaul capacity requirement at most $C$.

To this end, define a set function $f: 2^{\mathcal{K}} \rightarrow \mathbb{R}$ as follows:
\begin{equation*}
f\left(\mathcal{T}\right):= \min\left\{ C - \sum_{\ell\in \mathcal{L}}I(\mathbf{Y}_{\ell};\hat{\mathbf{Y}}_{\ell}|\mathbf{X}_{\mathcal{K}}), \enspace I\left(\mathbf{X}_{\mathcal{T}}; \hat{\mathbf{Y}}_{\mathcal{L}}|\mathbf{X}_{\mathcal{T}^c}\right)\right\},
\end{equation*}
for each $\mathcal{T} \subseteq \mathcal{K}$. It can be verified that the function $f$ is a submodular function (Appendix~\ref{append:submodular}, Lemma~\ref{lem:submodular-function}). By construction, $(R_1, R_2,\ldots, R_K)$ as defined by (\ref{def:JD-outerbound}) satisfies
\begin{equation*}
\displaystyle \sum_{k\in \mathcal{T}}R_{k}  \leq f\left(\mathcal{T}\right),
\end{equation*}
which is a submodular polyhedron associated with $f$. 

It follows by basic results in submodular optimization (Appendix \ref{append:submodular}, Proposition~\ref{prop:submodular-poly}) that, for a linear ordering $i_1 \prec i_2 \prec \cdots \prec i_K$ on the set $\mathcal{K}$, 
an extreme point of $\mathcal{R}^*_{JD,s}$ can be computed as follows
\begin{equation*}
\tilde{R}_{i_j} =  f\left(\{i_1,\ldots, i_j\}\right) - f\left(\{i_1,\ldots, i_{j-1}\}\right).
\end{equation*}
Furthermore, the extreme points of $\mathcal{R}^o_{JD}$ can be enumerated over all the orderings of $\mathcal{K}$.  Each ordering of $\mathcal{K}$ is analyzed in the same manner, hence for notational simplicity we only consider the natural ordering $i_j = j$ in the following proof.

By construction,
\begin{multline}\label{eqn:append-A-extrempoint}
\tilde{R}_{j} =  \min\left\{ C - \sum_{\ell\in \mathcal{L}}I(\mathbf{Y}_{\ell};\hat{\mathbf{Y}}_{\ell}|\mathbf{X}_{\mathcal{K}}), \enspace I\left(\mathbf{X}^j_{1}; \hat{\mathbf{Y}}_{\mathcal{L}}|\mathbf{X}^K_{j+1}\right)\right\}  \\
- \min\left\{ C - \sum_{\ell\in \mathcal{L}}I(\mathbf{Y}_{\ell};\hat{\mathbf{Y}}_{\ell}|\mathbf{X}_{\mathcal{K}}), \enspace I\left(\mathbf{X}^{j-1}_{1}; \hat{\mathbf{Y}}_{\mathcal{L}}|\mathbf{X}^K_{j}\right)\right\}.
\end{multline}
Due to the fact that $I\left(\mathbf{X}^j_{1}; \hat{\mathbf{Y}}_{\mathcal{L}}|\mathbf{X}^K_{j+1}\right) \geq I\left(\mathbf{X}^{j-1}_{1}; \hat{\mathbf{Y}}_{\mathcal{L}}|\mathbf{X}^K_{j}\right)$, for some product distribution $\prod_{k=1}^K p\left(\mathbf{x}_k\right)\prod_{\ell=1}^Lp(\hat{\mathbf{y}}_{\ell}|\mathbf{y}_{\ell})$, equation (\ref{eqn:append-A-extrempoint}) can yield two different results. Case 1: the first term $C - \sum_{\ell\in \mathcal{L}}I(\mathbf{Y}_{\ell};\hat{\mathbf{Y}}_{\ell}|\mathbf{X}_{\mathcal{K}})$ in the minima in equation (\ref{eqn:append-A-extrempoint}) is not active for any $j$; Case 2: the term $C - \sum_{\ell\in \mathcal{L}}I(\mathbf{Y}_{\ell};\hat{\mathbf{Y}}_{\ell}|\mathbf{X}_{\mathcal{K}})$ is active starting with some index $j$.

\begin{itemize}
\item Case 1 holds if  $C  \geq I\left(\mathbf{X}_{\mathcal{K}}; \hat{\mathbf{Y}}_{\mathcal{L}}\right) + \sum\limits_{\ell\in \mathcal{L}}I(\mathbf{Y}_{\ell};\hat{\mathbf{Y}}_{\ell}|\mathbf{X}_{\mathcal{K}})$.
    In this case the resulting extreme point $\mathbf{r}^1_{JD} = (\tilde{R}_1, \tilde{R}_2,\ldots, \tilde{R}_K, C)$ satisfies 
\begin{equation*}
\begin{cases}
\enspace \tilde{R}_j = I\left(\mathbf{X}_j; \hat{\mathbf{Y}}_{\mathcal{L}}|\mathbf{X}^K_{j+1}\right), \enspace \textrm{for} \enspace j=1,2,\ldots,K-1,\\
\enspace \tilde{R}_K = I\left(\mathbf{X}_K; \hat{\mathbf{Y}}_{\mathcal{L}}\right), \\
\enspace C = I\left(\mathbf{X}_{\mathcal{K}}; \hat{\mathbf{Y}}_{\mathcal{L}}\right) + \sum\limits_{\ell\in \mathcal{L}}I\left(\mathbf{Y}_{\ell};\hat{\mathbf{Y}}_{\ell}|\mathbf{X}_{\mathcal{K}}\right).
\end{cases}
\end{equation*}

Consider successive decoding with the decoding order $ \hat{\mathbf{Y}}_{\mathcal{L}} \rightarrow \mathbf{X}_{K}\rightarrow \cdots \rightarrow\mathbf{X}_{1}$. The extreme point $(R^*_1,\ldots,R^*_K, C^*) \in \mathcal{P}^*_{GSD,s}$ corresponding to this decoding order is
\begin{equation*}
\begin{cases}
\enspace \tilde{R}^*_j = I\left(\mathbf{X}_j; \hat{\mathbf{Y}}_{\mathcal{L}}|\mathbf{X}^K_{j+1}\right), \enspace \textrm{for} \enspace j=1,2,\ldots,K-1,\\
\enspace \tilde{R}^*_K = I\left(\mathbf{X}_K; \hat{\mathbf{Y}}_{\mathcal{L}}\right), \\
\enspace C^* = I(\mathbf{Y}_{\mathcal{L}};\hat{\mathbf{Y}}_{\mathcal{L}}).
\end{cases}
\end{equation*}
Following the Markov chain
\begin{equation*}
\hat{\mathbf{Y}}_{i} \leftrightarrow \mathbf{Y}_{i} \leftrightarrow
\mathbf{X}_{\mathcal{K}}
\leftrightarrow \mathbf{Y}_j \leftrightarrow \hat{\mathbf{Y}}_j, \quad \forall\; i\neq j,
\end{equation*}
it can be shown that
\begin{equation*}
 \sum\limits_{\ell\in \mathcal{L}}I(\mathbf{Y}_{\ell};\hat{\mathbf{Y}}_{\ell}|\mathbf{X}_{\mathcal{K}}) + I\left(\mathbf{X}_{\mathcal{K}}; \hat{\mathbf{Y}}_{\mathcal{L}}\right)= I(\mathbf{Y}_{\mathcal{L}};\hat{\mathbf{Y}}_{\mathcal{L}}). 
\end{equation*}
Clearly, $\mathbf{r}^1_{JD}$ can be achieved by the decoding order of $ \hat{\mathbf{Y}}_{\mathcal{L}} \rightarrow \mathbf{X}_{K}\rightarrow \cdots \rightarrow\mathbf{X}_{1}$. Thus, $\mathbf{r}^1_{JD}$ is dominated by a point in $\mathcal{P}^*_{GSD,s}$.

\item Case 2 holds if $C \leq I\left(\mathbf{X}_{\mathcal{K}}; \hat{\mathbf{Y}}_{\mathcal{L}}\right) + \sum\limits_{\ell\in \mathcal{L}}I(\mathbf{Y}_{\ell};\hat{\mathbf{Y}}_{\ell}|\mathbf{X}_{\mathcal{K}})$. We let $\mathbf{X}^i_{j} = \emptyset$ for $i<j$, and assume that
\begin{equation*}
I\left(\mathbf{X}^{j-1}_{1}; \hat{\mathbf{Y}}_{\mathcal{L}}|\mathbf{X}^K_{j}\right) \leq C - \sum\limits_{\ell\in \mathcal{L}}I(\mathbf{Y}_{\ell};\hat{\mathbf{Y}}_{\ell}|\mathbf{X}_{\mathcal{K}})
\end{equation*}
and
\begin{equation*}
C - \sum\limits_{\ell\in \mathcal{L}}I(\mathbf{Y}_{\ell};\hat{\mathbf{Y}}_{\ell}|\mathbf{X}_{\mathcal{K}}) \leq I\left(\mathbf{X}^j_{1}; \hat{\mathbf{Y}}_{\mathcal{L}}|\mathbf{X}^K_{j+1}\right)
\end{equation*}
for some $1\leq j\leq K$. The resulting extreme point $\mathbf{r}^2_{JD} = (\tilde{R}_1, \tilde{R}_2,\ldots, \tilde{R}_K, C)$ satisfies 
\begin{equation*}
\begin{cases}
\tilde{R}_i = I\left(\mathbf{X}_i; \hat{\mathbf{Y}}_{\mathcal{L}}|\mathbf{X}^K_{i+1}\right),  \quad \textrm{for} \enspace i<j,\\
\tilde{R}_i = \left[C - \sum\limits_{\ell\in \mathcal{L}}I(\mathbf{Y}_{\ell};\hat{\mathbf{Y}}_{\ell}|\mathbf{X}_{\mathcal{K}}) - I\left(\mathbf{X}^{j-1}_{1}; \hat{\mathbf{Y}}_{\mathcal{L}}|\mathbf{X}^K_{i}\right)\right]^+, \\
\hspace{6cm} \enspace \textrm{for} \enspace i=j,\\
\tilde{R}_i = 0, \quad \textrm{for} \enspace i>j, \\
C  = I\left(\mathbf{X}^j_{1}; \hat{\mathbf{Y}}_{\mathcal{L}}|\mathbf{X}^K_{j+1}\right) + \sum\limits_{\ell\in \mathcal{L}}I(\mathbf{Y}_{\ell};\hat{\mathbf{Y}}_{\ell}|\mathbf{X}_{\mathcal{K}}).
\end{cases}
\end{equation*}
where $[\cdot]^+$ means $\max\{\cdot, 0\}$. Note that users with index $i>j$ are inactive, and are essentially removed from the network. In this case, the rate-fronthaul tuple does not correspond to a specific corner point obtained with a specific generalized successive decoding order, but that it lies on the convex-hull of
two corner points of two different generalized successive decoding orders. To obtain a visualization on Case 2, the rate-fronthaul region for a two-user C-RAN model under a fixed joint distribution $p\left(\mathbf{x}_1, \mathbf{x}_2, \mathbf{y}_{1}, \mathbf{y}_{2}, \hat{\mathbf{y}}_{1}, \hat{\mathbf{y}}_{2} \right)$ is illustrated in Fig.~\ref{fig:append-A-fig}. In the case of $K=j=2$, it is shown that the rate-fronthaul tuple $\mathbf{r}^2_{JD}$ lies on the convex-hull of two corner points $\mathbf{r}^{(1)}$ and $\mathbf{r}^{(2)}$.


\begin{figure} [t]
    \centering
    \begin{overpic}[width=0.45\textwidth]{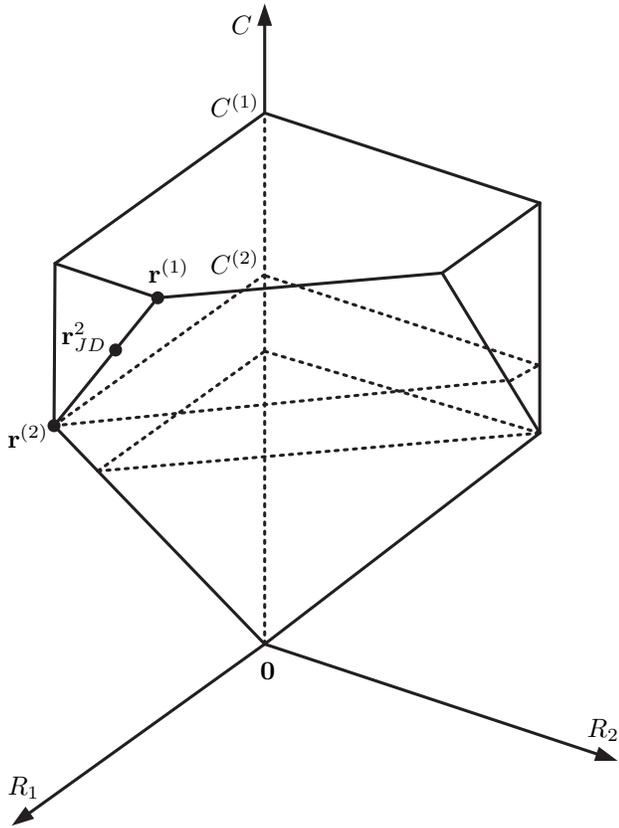}
    \put(27,96){$C$}
    \put(24.5,86){$C^{(1)}$}
    \put(24.5,67){$C^{(2)}$}
    \put(0,4){$R_1$}
    \put(70,11){$R_2$}
    \put(30.5,18){$\mathbf{0}$}
    \put(17,66){$\mathbf{r}^{(1)}$}
    \put(0,46){$\mathbf{r}^{(2)}$}
    \put(6.5,58.5){$\mathbf{r}^2_{JD}$}
    \end{overpic}
    \caption{An illustration of the rate-fronthaul tuple in Case 2 in Appendix \ref{append:proof-GSD=JD} with a two-user C-RAN model under a fixed joint distribution $p\left(\mathbf{x}_1, \mathbf{x}_2, \mathbf{y}_{1}, \mathbf{y}_{2}, \hat{\mathbf{y}}_{1}, \hat{\mathbf{y}}_{2} \right)$.}
    \label{fig:append-A-fig}
\end{figure}

To prove the statement mathematically, we consider generalized successive decoding with the following two different decoding orders:
(i) Decoding order 1 satisfies
\begin{equation*}
\mathbf{X}_{K}\rightarrow\ldots\rightarrow\mathbf{X}_{j+1} \rightarrow \hat{\mathbf{Y}}_{\mathcal{L}} \rightarrow \mathbf{X}_j \rightarrow\ldots\rightarrow\mathbf{X}_1.
\end{equation*}
The extreme point $\mathbf{r}^{(1)}_{GSD}=(R^{(1)}_1,\ldots,R^{(1)}_K, C^{(1)})$ of $\mathcal{P}^*_{GSD,s}$ corresponding to Decoding order 1 satisfies
\begin{equation*}
\begin{cases}
\enspace R^{(1)}_i = I\left(\mathbf{X}_i; \hat{\mathbf{Y}}_{\mathcal{L}}|\mathbf{X}^K_{i+1}\right), & \enspace \textrm{for} \enspace i\leq j,\\
\enspace R^{(1)}_i = 0, &\enspace \textrm{for} \enspace i>j, \\
\enspace C^{(1)} = I\left(\mathbf{Y}_{\mathcal{L}};\hat{\mathbf{Y}}_{\mathcal{L}}|\mathbf{X}^K_{j+1}\right).
\end{cases}
\end{equation*}
where $C^{(1)}$ represents the required fronthaul capacity in order to achieve the above rate tuple $(R^{(1)}_1,\ldots,R^{(1)}_K)$ with decoding order 1.

(ii) Decoding order 2 is
\begin{equation*}
\mathbf{X}_{K}\rightarrow\ldots\rightarrow\mathbf{X}_j \rightarrow \hat{\mathbf{Y}}_{\mathcal{L}} \rightarrow \mathbf{X}_{j-1} \rightarrow\ldots\rightarrow\mathbf{X}_1.
\end{equation*}
The extreme point $\mathbf{r}^{(2)}_{GSD}=(R^{(2)}_1,\ldots,R^{(2)}_K, C^{(2)})$ of $\mathcal{P}^*_{GSD,s}$ corresponding to Decoding order 2 satisfies
\begin{equation*}
\begin{cases}
\enspace R^{(2)}_i = I\left(\mathbf{X}_i; \hat{\mathbf{Y}}_{\mathcal{L}}|\mathbf{X}^K_{i+1}\right), & \enspace \textrm{for} \enspace i< j,\\
\enspace R^{(2)}_i = 0, &\enspace \textrm{for} \enspace i\geq j, \\
\enspace C^{(2)} = I\left(\mathbf{Y}_{\mathcal{L}};\hat{\mathbf{Y}}_{\mathcal{L}}|\mathbf{X}^K_{j}\right).
\end{cases}
\end{equation*}
where $C^{(1)}$ represents the required fronthaul capacity in order to achieve the above rate tuple $(R^{(2)}_1,\ldots,R^{(2)}_K)$ with decoding order 2. Observe that the rate tuples $(R^{(1)}_1,\ldots,R^{(1)}_K)$ and $(R^{(2)}_1,\ldots,R^{(2)}_K)$ given by above two decoding orders different at only the $j$th component, where $R^{(1)}_j = I\left(\mathbf{X}_j; \hat{\mathbf{Y}}_{\mathcal{L}}|\mathbf{X}^K_{j+1}\right)$ and $R^{(2)}_j = 0$ and $R^{(1)}_i = R^{(2)}_i = \tilde{R}_i$ for all $i<j$. Now choose a  parameter $\theta$ such that
\begin{equation}
\label{eqn:theta}
\theta  = \frac{ C - \sum\limits_{\ell\in \mathcal{L}}I(\mathbf{Y}_{\ell};\hat{\mathbf{Y}}_{\ell}|\mathbf{X}_{\mathcal{K}}) - I\left(\mathbf{X}^{j-1}_{1}; \hat{\mathbf{Y}}_{\mathcal{L}}|\mathbf{X}^K_{j}\right)}{I\left(\mathbf{X}_j; \hat{\mathbf{Y}}_{\mathcal{L}}|\mathbf{X}^K_{j+1}\right)}.
\end{equation}
Following the Markov chain $\mathbf{X}_{\mathcal{K}} \leftrightarrow \mathbf{Y}_{\mathcal{L}}\leftrightarrow \hat{\mathbf{Y}}_{\mathcal{L}}$, we have the following identity,
\begin{align*}
1-\theta
& = \frac{I\left(\mathbf{Y}_{\mathcal{L}}; \hat{\mathbf{Y}}_{\mathcal{L}}|\mathbf{X}^K_{j+1}\right) - C }{I\left(\mathbf{X}_j; \hat{\mathbf{Y}}_{\mathcal{L}}|\mathbf{X}^K_{j+1}\right)}.
\end{align*}

Consider the following point:
$\mathbf{r}^{\theta}_{GSD} = \theta\mathbf{r}^{(1)}_{GSD} + (1-\theta)\mathbf{r}^{(2)}_{GSD}$, which is in $\mathcal{P}^*_{GSD,s}$. The corresponding sum fronthaul requirement is given by
\begin{align}
\label{eqn:front-require-GSD}
& \theta C^{(1)} + (1-\theta)C^{(2)} \nonumber \\
& = \theta I\left(\mathbf{Y}_{\mathcal{L}};\hat{\mathbf{Y}}_{\mathcal{L}}|\mathbf{X}^K_{j+1}\right) + (1-\theta)I\left(\mathbf{Y}_{\mathcal{L}};\hat{\mathbf{Y}}_{\mathcal{L}}|\mathbf{X}^K_{j}\right) \nonumber \\
& = C \times \frac{I\left(\mathbf{Y}_{\mathcal{L}};\hat{\mathbf{Y}}_{\mathcal{L}}|\mathbf{X}^K_{j+1}\right) - I\left(\mathbf{Y}_{\mathcal{L}};\hat{\mathbf{Y}}_{\mathcal{L}}|\mathbf{X}^K_{j}\right)}{I\left(\mathbf{X}_j; \hat{\mathbf{Y}}_{\mathcal{L}}|\mathbf{X}^K_{j+1}\right)} \nonumber \\
& \quad +  \frac{I\left(\mathbf{Y}_{\mathcal{L}};\hat{\mathbf{Y}}_{\mathcal{L}}|\mathbf{X}^K_{j+1}\right)}{I\left(\mathbf{X}_j; \hat{\mathbf{Y}}_{\mathcal{L}}|\mathbf{X}^K_{j+1}\right)}  \times
\left[I\left(\mathbf{Y}_{\mathcal{L}};\hat{\mathbf{Y}}_{\mathcal{L}}|\mathbf{X}^K_{j}\right) \right. \nonumber\\
& \qquad \qquad \quad \left.- I\left(\mathbf{Y}_{\mathcal{L}}; \hat{\mathbf{Y}}_{\mathcal{L}}|\mathbf{X}^K_{1}\right)
 -I\left(\mathbf{X}^{j-1}_{1}; \hat{\mathbf{Y}}_{\mathcal{L}}|\mathbf{X}^K_{j}\right)\right] \nonumber\\
& \overset{(c)}{=} C \times \frac{I\left(\mathbf{Y}_{\mathcal{L}};\hat{\mathbf{Y}}_{\mathcal{L}}|\mathbf{X}^K_{j+1}\right) - I\left(\mathbf{Y}_{\mathcal{L}};\hat{\mathbf{Y}}_{\mathcal{L}}|\mathbf{X}^K_{j}\right)}{I\left(\mathbf{X}_j; \hat{\mathbf{Y}}_{\mathcal{L}}|\mathbf{X}^K_{j+1}\right)} \nonumber \\
& \quad +  \frac{I\left(\mathbf{Y}_{\mathcal{L}};\hat{\mathbf{Y}}_{\mathcal{L}}|\mathbf{X}^K_{j+1}\right)}{I\left(\mathbf{X}_j; \hat{\mathbf{Y}}_{\mathcal{L}}|\mathbf{X}^K_{j+1}\right)} \times
\left[I\left(\mathbf{X}^{j-1}_{1}, \mathbf{Y}_{\mathcal{L}};\hat{\mathbf{Y}}_{\mathcal{L}}|\mathbf{X}^K_{j}\right) \right. \nonumber \\
& \qquad \qquad \quad \left.- I\left(\mathbf{Y}_{\mathcal{L}}; \hat{\mathbf{Y}}_{\mathcal{L}}|\mathbf{X}^K_{1}\right)
 -I\left(\mathbf{X}^{j-1}_{1}; \hat{\mathbf{Y}}_{\mathcal{L}}|\mathbf{X}^K_{j}\right)\right] \nonumber\\
& \overset{(d)}{\leq} C \times \frac{I\left(\mathbf{X}_j,\mathbf{Y}_{\mathcal{L}};\hat{\mathbf{Y}}_{\mathcal{L}}|\mathbf{X}^K_{j+1}\right) - I\left(\mathbf{Y}_{\mathcal{L}};\hat{\mathbf{Y}}_{\mathcal{L}}|\mathbf{X}^K_{j}\right)}{I\left(\mathbf{X}_j; \hat{\mathbf{Y}}_{\mathcal{L}}|\mathbf{X}^K_{j+1}\right)} \nonumber\\
& = C,
\end{align}
where the equality $(c)$ follows from the fact that $I\left(\mathbf{X}^{j-1}_{1},\mathbf{Y}_{\mathcal{L}};\hat{\mathbf{Y}}_{\mathcal{L}}|\mathbf{X}^K_{j}\right) = I\left(\mathbf{Y}_{\mathcal{L}};\hat{\mathbf{Y}}_{\mathcal{L}}|\mathbf{X}^K_{j}\right)$ due to Markov chain $\mathbf{X}_{\mathcal{K}} \leftrightarrow \mathbf{Y}_{\mathcal{L}}\leftrightarrow \hat{\mathbf{Y}}_{\mathcal{L}}$, and inequality $(d)$ follows from the fact that $I\left(\mathbf{Y}_{\mathcal{L}};\hat{\mathbf{Y}}_{\mathcal{L}}|\mathbf{X}^K_{j+1}\right) \leq I\left(\mathbf{X}_j,\mathbf{Y}_{\mathcal{L}};\hat{\mathbf{Y}}_{\mathcal{L}}|\mathbf{X}^K_{j+1}\right)$.
Thus, we have that $\mathbf{r}^2_{JD}$ is dominated by some point lying on line segment between $\mathbf{r}^{(1)}_{GSD}$ and $\mathbf{r}^{(2)}_{GSD}$, which lies in $\mathcal{P}^*_{GSD,s}$.
\end{itemize}

Therefore, for every extreme point $(\tilde{R}_1,\ldots, \tilde{R}_K) $ of $\mathcal{R}^o_{JD}$, the point $(\tilde{R}_1,\ldots, \tilde{R}_K, C) $ lies in $\mathcal{P}^*_{GSD,s}$. This completes the proof.

\section{Submodular Functions}
\label{append:submodular}
In this appendix, we review some basic results in submodular optimization used proving Theorem~\ref{thm:GSD=JD-SumFront} and Theorem~\ref{thm:opt-VMAC}. We tailor our statements toward submodularity and supermodularity, which are used in the proofs.

We begin with the definition of submodular function.
\begin{defn}
Let $ \mathcal{D}= \{1, \ldots, n\}$ be a finite set. A set function $f: 2^{\mathcal{D}} \rightarrow \mathbb{R}$ is submodular if for all $\mathcal{S}, \mathcal{T} \subseteq \mathcal{D}$,
\begin{equation}
\label{def:submodular-function}
f(\mathcal{S}) + f(\mathcal{T}) \geq f(\mathcal{S}\cup \mathcal{T}) + f(\mathcal{S}\cap \mathcal{T}).
\end{equation}
\end{defn}

\begin{defn}
Let $ \mathcal{E}= \{1, \ldots, m\}$ be a finite set. A set function $g: 2^{\mathcal{E}} \rightarrow \mathbb{R}$ is supermodular if for all $\mathcal{S}, \mathcal{T} \subseteq \mathcal{E}$,
\begin{equation}
\label{def:supermodular-function}
g(\mathcal{S}) + g(\mathcal{T}) \leq g(\mathcal{S}\cup \mathcal{T}) + g(\mathcal{S}\cap \mathcal{T}).
\end{equation}
\end{defn}

If the function $f$ is submodular, we call a polyhedron defined by
\begin{equation}
\mathcal{P}(f) = \left\{(x_1,\ldots,x_n)\in \mathbb{R}^n : \sum_{i\in \mathcal{S}} x_i \leq f(\mathcal{S}),\enspace \forall \; \mathcal{S}\subseteq \mathcal{D}\right\}
\end{equation}
the submodular polyhedron associated with the submodular function $f$. Similarly, we define the supermodular polyhedron $\mathcal{P}(g) $ to be the set of $(x_1,\ldots,x_n)\in \mathbb{R}^n$ satisfying
\begin{equation}
 \sum_{i\in \mathcal{T}} x_i \geq g(\mathcal{T}),\enspace \forall \; \mathcal{T}\subseteq \mathcal{E}.
\end{equation}
We say a point in $\mathcal{P}(f)$ is an extreme point if it cannot be expressed as a convex combination of the other two points in $\mathcal{P}(f)$.

One important property of submodular polyhedron is that all the extreme points can be enumerated through solving a linear optimization. The following proposition provides an algorithm that enumerates the extreme points of $\mathcal{P}(f)$.
\begin{prop}[{\cite{fujishige2005submodular}~\cite{iwata2008submodular}}]
\label{prop:submodular-poly}
For a linear ordering  $i_1 \prec i_2 \prec \cdots \prec i_n$ of the elements in $\mathcal{D}$, Algorithm~\ref{alg:submodular-poly}  returns an extreme point $(v_1, \ldots, v_n)$ of $\mathcal{P}(f)$. Moreover, all extreme points of $\mathcal{P}(f)$ can be enumerated by considering all linear orderings of the
elements of $\mathcal{D}$.
\begin{algorithm}[!ht]
\caption{Greedy Algorithm for Submodular Polyhedron}
\begin{algorithmic}[1]
\STATE \textbf{comment}: Returns extreme point $(v_1, \ldots, v_n)$ of $\mathcal{P}(f)$ with the ordering $\prec$.
\FOR {$j=1,\ldots,n$}
\STATE Set $v_j = f\left(\{i_1,i_2,\ldots, i_j\}\right)- f\left(\{i_1,i_2,\ldots, i_{j-1}\}\right)$.
\ENDFOR
\RETURN {$(v_1, \ldots, v_n)$}
\end{algorithmic}
\label{alg:submodular-poly}
\end{algorithm}
\end{prop}

Proposition~\ref{prop:submodular-poly} is the key tool we employ to prove Theorem~\ref{thm:GSD=JD-SumFront} and Theorem~\ref{thm:opt-VMAC}. In order to apply this proposition, we require the following lemmas,
\begin{lemma}
\label{lem:submodular-function}
For any joint distribution $\prod_{k=1}^Kp\left(\mathbf{x}_k\right)\prod_{\ell=1}^Lp\left(\mathbf{y}_{\ell}|\mathbf{x}_{1}^K\right)\prod_{\ell=1}^Lp(\hat{\mathbf{y}}_{\ell}|\mathbf{y}_{\ell})$ and fixed $C\in \mathbb{R}$, the set function $f: 2^{\mathcal{K}} \rightarrow \mathbb{R}$ defined as follows
\begin{equation*}
f\left(\mathcal{T}\right):= \min\left\{ C - \sum_{\ell\in \mathcal{L}}I(\mathbf{Y}_{\ell};\hat{\mathbf{Y}}_{\ell}|\mathbf{X}_{\mathcal{K}}), \enspace I\left(\mathbf{X}_{\mathcal{T}}; \hat{\mathbf{Y}}_{\mathcal{L}}|\mathbf{X}_{\mathcal{T}^c}\right)\right\}
\end{equation*}
is submodular.
\end{lemma}

\begin{IEEEproof}
Define a set function $ f'\left(\mathcal{T}\right)= I\left(\mathbf{X}_{\mathcal{T}}; \hat{\mathbf{Y}}_{\mathcal{L}}|\mathbf{X}_{\mathcal{T}^c}\right)$. By definition, it can be verified that function $f'$ is submodular~\cite{zhang2007successive}. Under fixed sum fronthaul capacity $C$ and conditional distribution $\prod_{\ell=1}^Lp_{\hat{\mathbf{Y}}_{\ell}|\mathbf{Y}_{\ell}}$, the expression $C - \sum_{\ell\in \mathcal{L}}I(\mathbf{Y}_{\ell};\hat{\mathbf{Y}}_{\ell}|\mathbf{X}_{\mathcal{K}})$ is a constant. Let $C'= C - \sum_{\ell\in \mathcal{L}}I(\mathbf{Y}_{\ell};\hat{\mathbf{Y}}_{\ell}|\mathbf{X}_{\mathcal{K}})$. Now the problem reduces to show that $f\left(\mathcal{T}\right)= \min\left\{C', f'\left(\mathcal{T}\right)\right)$ is submodular.

Next, observe that $f'$ is monotonically increasing, i.e., if $\mathcal{S}\subset \mathcal{T}$, then $f'(\mathcal{S})\leq f'(\mathcal{T})$. Thus, fixing $\mathcal{S}, \mathcal{T}\subseteq \mathcal{K}$, we can assume without loss of generality that
\begin{equation*}
f'(\mathcal{S}\cap \mathcal{T}) \leq f'(\mathcal{S})\leq f'(\mathcal{T})\leq f'(\mathcal{S}\cup \mathcal{T})
\end{equation*}
If $C'\leq f'(\mathcal{S}\cap \mathcal{T})$, then $f(\mathcal{S})= f(\mathcal{T}) =  f(\mathcal{S}\cap \mathcal{T}) = f(\mathcal{T})\leq f'(\mathcal{S}\cup \mathcal{T})= C'$. Clearly,  $f$ is then submodular. On the other hand, if $C'\geq f'(\mathcal{S}\cup \mathcal{T})$, then $f(\mathcal{S}) = f'(\mathcal{S})$, $f(\mathcal{T}) = f'(\mathcal{T})$, $f(\mathcal{S}\cap \mathcal{T}) = f'(\mathcal{S}\cap \mathcal{T})$, and $f(\mathcal{S}\cup \mathcal{T})= f'(\mathcal{S}\cup \mathcal{T})$, $f$ is also submodular. Thus, it suffices to check the following three cases:
\begin{itemize}
\item Case 1: $f'(\mathcal{S}\cap \mathcal{T}) \leq C' \leq f'(\mathcal{S})\leq f'(\mathcal{T})\leq f'(\mathcal{S}\cup \mathcal{T})$.

By definition of function $f$, we have
\begin{equation*}
f(\mathcal{S}) + f(\mathcal{T}) \geq  C' + f'(\mathcal{S}\cap \mathcal{T})
= f(\mathcal{S}\cup \mathcal{T}) + f(\mathcal{S}\cap \mathcal{T}).
\end{equation*}

\item Case 2: $f'(\mathcal{S}\cap \mathcal{T})  \leq f'(\mathcal{S}) \leq C' \leq f'(\mathcal{T})\leq f'(\mathcal{S}\cup \mathcal{T})$.

Since $f'$ is monotonically increasing, we have
\begin{eqnarray*}
f(\mathcal{S}) + f(\mathcal{T})  =   f'(\mathcal{S}) + C' &\geq&  f'(\mathcal{S}\cap \mathcal{T}) + f(\mathcal{S}\cup \mathcal{T}) \\
&= & f(\mathcal{S}\cap \mathcal{T}) + f(\mathcal{S}\cup \mathcal{T}) .
\end{eqnarray*}

\item Case 3: $f'(\mathcal{S}\cap \mathcal{T})  \leq f'(\mathcal{S})  \leq f'(\mathcal{T}) \leq C' \leq f'(\mathcal{S}\cup \mathcal{T})$.

In this case, the submodularity of $f'$ and the fact of $f'\leq f$ imply that
\begin{eqnarray*}
f(\mathcal{S}) + f(\mathcal{T})  &=&  f'(\mathcal{S})  + f'(\mathcal{T}) \\
&\geq & f'(\mathcal{S}\cap \mathcal{T}) + f'(\mathcal{S}\cup \mathcal{T}) \\
&\geq & f(\mathcal{S}\cap \mathcal{T}) + f(\mathcal{S}\cup \mathcal{T}) .
\end{eqnarray*}
\end{itemize}
Hence, $f = \min\{C', f'\}$ is submodular, which completes the proof of Lemma~\ref{lem:submodular-function}.
\end{IEEEproof}

\begin{lemma}
\label{lem:supermodular-function}
For any joint distribution $\prod_{k=1}^Kp\left(\mathbf{x}_k\right)\prod_{\ell=1}^Lp\left(\mathbf{y}_{\ell}|\mathbf{x}_{1}^K\right)\prod_{\ell=1}^Lp(\hat{\mathbf{y}}_{\ell}|\mathbf{y}_{\ell})$  and fixed $R \in \mathbb{R}$, define the set function $g: 2^{\mathcal{L}} \rightarrow \mathbb{R}$ as:
\begin{equation*}
g\left(\mathcal{S}\right):= R + \sum_{\ell\in \mathcal{S}}I\left(\mathbf{Y}_{\ell};\hat{\mathbf{Y}}_{\ell}|\mathbf{X}_{\mathcal{K}}\right) - I\left(\mathbf{X}_{\mathcal{K}}; \hat{\mathbf{Y}}_{\mathcal{S}^c}\right),
\end{equation*}
and the corresponding non-negative set function $g^+:2^{\mathcal{L}} \rightarrow \mathbb{R}_+$ as $g^+ = \max\{g, 0\}$. The functions $g$ and $g^+$ are supermodular.
\end{lemma}

\begin{IEEEproof}
We first prove that the set function $g'\left(\mathcal{T}\right)= I\left(\mathbf{X}_{\mathcal{K}}; \hat{\mathbf{Y}}_{\mathcal{T}}\right)$ is submodular. To this end, we evaluate
\begin{eqnarray*}
& & g'\left(\mathcal{T}\cap \mathcal{S}\right)  + g'\left(\mathcal{T}\cup\mathcal{S}\right) \\
&=& I\left(\mathbf{X}_{\mathcal{K}}; \hat{\mathbf{Y}}_{\mathcal{T}\cup \mathcal{S}}\right) + I\left(\mathbf{X}_{\mathcal{K}}; \hat{\mathbf{Y}}_{\mathcal{T}\cap \mathcal{S}}\right) \\
&=& I\left(\mathbf{X}_{\mathcal{K}}; \hat{\mathbf{Y}}_{\mathcal{S}}, \hat{\mathbf{Y}}_{\mathcal{S}^c \cap \mathcal{T}}\right) + I\left(\mathbf{X}_{\mathcal{K}}; \hat{\mathbf{Y}}_{\mathcal{T}\cap \mathcal{S}}\right) \\
&=& g'\left(\mathcal{S}\right) + g'\left(\mathcal{T}\right) +  I\left(\mathbf{X}_{\mathcal{K}}; \hat{\mathbf{Y}}_{\mathcal{S}^c \cap \mathcal{T}}| \hat{\mathbf{Y}}_{\mathcal{S}}\right)  \\
& & \qquad \qquad \qquad \qquad \enspace - I\left(\mathbf{X}_{\mathcal{K}};  \hat{\mathbf{Y}}_{\mathcal{S}^c \cap \mathcal{T}}| \hat{\mathbf{Y}}_{\mathcal{T}\cap \mathcal{S}}\right).
\end{eqnarray*}

Furthermore,
\begin{eqnarray*}
&& I\left(\mathbf{X}_{\mathcal{K}}; \hat{\mathbf{Y}}_{\mathcal{S}^c \cap \mathcal{T}}| \hat{\mathbf{Y}}_{\mathcal{S}}\right)  - I\left(\mathbf{X}_{\mathcal{K}};  \hat{\mathbf{Y}}_{\mathcal{S}^c \cap \mathcal{T}}| \hat{\mathbf{Y}}_{\mathcal{T}\cap \mathcal{S}}\right) \\
&=& h\left(\hat{\mathbf{Y}}_{\mathcal{S}^c \cap \mathcal{T}}| \hat{\mathbf{Y}}_{\mathcal{S}}\right) - h\left(\hat{\mathbf{Y}}_{\mathcal{S}^c \cap \mathcal{T}}| \hat{\mathbf{Y}}_{\mathcal{S}}, \mathbf{X}_{\mathcal{K}}\right) \\
& & \quad - h\left(\hat{\mathbf{Y}}_{\mathcal{S}^c \cap \mathcal{T}}| \hat{\mathbf{Y}}_{\mathcal{T}\cap \mathcal{S}}\right) + h\left(\hat{\mathbf{Y}}_{\mathcal{S}^c \cap \mathcal{T}}| \hat{\mathbf{Y}}_{\mathcal{T}\cap \mathcal{S}}, \mathbf{X}_{\mathcal{K}}\right) \\
&=& h\left(\hat{\mathbf{Y}}_{\mathcal{S}^c \cap \mathcal{T}}| \hat{\mathbf{Y}}_{\mathcal{S}}\right) - h\left(\hat{\mathbf{Y}}_{\mathcal{S}^c \cap \mathcal{T}}| \hat{\mathbf{Y}}_{\mathcal{S}\cap\mathrm{T}}\right) \\
&\leq& 0.
\end{eqnarray*}

Therefore, $g'\left(\mathcal{T}\cap \mathcal{S}\right)  + g'\left(\mathcal{T}\cup\mathcal{S}\right) \leq  g'\left(\mathcal{S}\right) + g'\left(\mathcal{T}\right)$, which proves that $g'$ is submodular.

In the following, we prove that $g$ is supermodular. Evaluate $g(\mathcal{S}) + g(\mathcal{T})$ as
\begin{eqnarray*}
&& g(\mathcal{S}) + g(\mathcal{T}) \\
&=& 2R + \sum_{\ell\in \mathcal{S}}I\left(\mathbf{Y}_{\ell};\hat{\mathbf{Y}}_{\ell}|\mathbf{X}_{\mathcal{K}}\right) + \sum_{\ell\in \mathcal{T}}I\left(\mathbf{Y}_{\ell};\hat{\mathbf{Y}}_{\ell}|\mathbf{X}_{\mathcal{K}}\right) \\
& & \quad - I\left(\mathbf{X}_{\mathcal{K}}; \hat{\mathbf{Y}}_{\mathcal{S}^c}\right) - I\left(\mathbf{X}_{\mathcal{K}}; \hat{\mathbf{Y}}_{\mathcal{T}^c}\right) \\
&\overset{(e)}{\leq} & 2R + \sum_{\ell\in \mathcal{S}\cup \mathcal{T}}I\left(\mathbf{Y}_{\ell};\hat{\mathbf{Y}}_{\ell}|\mathbf{X}_{\mathcal{K}}\right) + \sum_{\ell\in \mathcal{S}\cap \mathcal{T}}I\left(\mathbf{Y}_{\ell};\hat{\mathbf{Y}}_{\ell}|\mathbf{X}_{\mathcal{K}}\right) \\
& & \quad - I\left(\mathbf{X}_{\mathcal{K}}; \hat{\mathbf{Y}}_{(\mathcal{S}\cap\mathcal{T})^c}\right) - I\left(\mathbf{X}_{\mathcal{K}}; \hat{\mathbf{Y}}_{(\mathcal{S}\cup\mathcal{T})^c}\right) \\
&=& g(\mathcal{S}\cap \mathcal{T}) + g(\mathcal{S}\cup \mathcal{T}),
\end{eqnarray*}
where inequality (e) follows from the fact that $g'\left(\mathcal{T}\right)= I\left(\mathbf{X}_{\mathcal{K}}; \hat{\mathbf{Y}}_{\mathcal{T}}\right)$ is a submodular function.

Therefore, we show that $g$ is supermodular. Following the result of \cite[Lemma 6]{Courtade14}, it can be shown that $g^+ = \max\{g, 0\} $ is also supermodular.
\end{IEEEproof}

\section{Optimality of Successive Decoding for Maximizing Sum Rate}
\label{append:Opt-VMAC}

Similar to the proof of Theorem~\ref{thm:GSD=JD-SumFront}, Theorem~\ref{thm:opt-VMAC} can also be proven using submodular optimization. In the following, we consider the region $(R, C_1, \ldots, C_L)$, and prove that joint decoding and successive decoding achieve the same maximum rate using the properties of submodular optimization.

\begin{defn}
Define $\mathcal{P}^s_{JD}$ to be the closure of the convex hull of all $(R, C_1, \ldots, C_L)$ satisfying
\begin{equation}\label{eqn:SumRate-JD-poly}
\displaystyle R <  \sum_{\ell\in \mathcal{S}}\left[C_{\ell} - I\left(\mathbf{Y}_{\ell};\hat{\mathbf{Y}}_{\ell}|\mathbf{X}_{\mathcal{K}}\right) \right] + I\left(\mathbf{X}_{\mathcal{K}}; \hat{\mathbf{Y}}_{\mathcal{S}^c}\right), \enspace \forall \; \mathcal{S}\subseteq \mathcal{L}, 
\end{equation}
for some product distribution $\prod_{k=1}^Kp\left(\mathbf{x}_k\right)\prod_{\ell=1}^Lp(\hat{\mathbf{y}}_{\ell}|\mathbf{y}_{\ell})$.
\end{defn}

\begin{defn}
Define $\mathcal{P}^s_{SD}$ to be the closure of the convex hull all $(R, C_1, \ldots, C_L)$ satisfying
\begin{equation}\label{eqn:SumRate-SD-poly}
\begin{cases}
\enspace \displaystyle R < I\left(\mathbf{X}_{\mathcal{K}}; \hat{\mathbf{Y}}_{\mathcal{L}}\right), \\
\enspace \displaystyle \sum_{\ell\in \mathcal{S}}C_{\ell} > I\left(\mathbf{Y}_{\mathcal{S}}; \hat{\mathbf{Y}}_{\mathcal{S}}| \hat{\mathbf{Y}}_{\mathcal{S}^c}\right), \quad \forall \; \mathcal{S} \subseteq \mathcal{L}
\end{cases}
\end{equation}
for some product distribution $\prod_{k=1}^Kp\left(\mathbf{x}_k\right)\prod_{\ell=1}^Lp(\hat{\mathbf{y}}_{\ell}|\mathbf{y}_{\ell})$.
\end{defn}

Note that $\mathcal{P}^s_{JD}$ represents the sum-rate and fronthaul-capacity region of joint decoding. All the partial sums over $\mathcal{S}$ in (\ref{eqn:SumRate-JD-poly}) can be strictly attained with equality depending on the values of the fronthaul capacities $C_{\ell}$ for $\ell=1,\ldots,L$ and the sum rate $R$. Similarly, $\mathcal{P}^s_{SD}$ corresponds to the region of successive decoding. For fixed product distribution $\prod_{k=1}^Kp\left(\mathbf{x}_k\right)\prod_{\ell=1}^Lp(\hat{\mathbf{y}}_{\ell}|\mathbf{y}_{\ell})$, we say a point $(R, C_1, \ldots, C_L)$ is dominated by a point $(R', C'_1, \ldots, C'_L)$ in $\mathcal{P}^s_{SD}$ if $C'_{\ell} \leq C_{\ell}$ for $\ell = 1,\ldots,L$ and $R' \geq R$.

Clearly, the maximum sum rate achieved by joint decoding is always larger or equal to that achieved by successive decoding, i.e., $R^*_{JD, SUM} \geq R^*_{SD, SUM}$. To show $R^*_{JD, SUM} = R^*_{SD, SUM}$, it remains to show that $R^*_{JD, SUM} \leq R^*_{SD, SUM}$. For any given product distribution  $\prod_{k=1}^Kp\left(\mathbf{x}_k\right)\prod_{\ell=1}^Lp(\hat{\mathbf{y}}_{\ell}|\mathbf{y}_{\ell})$ and joint decoding sum rate $R_{JD}$, define $\mathcal{P}_{C} \subset \mathbb{R}^L_+$ to be the set of $(C_1,\ldots, C_L)$ such that
\begin{equation}
\sum_{\ell\in \mathcal{S}}C_{\ell}  \geq \left[R_{JD} + \sum_{\ell\in \mathcal{S}}I\left(\mathbf{Y}_{\ell};\hat{\mathbf{Y}}_{\ell}|\mathbf{X}_{\mathcal{K}}\right) - I\left(\mathbf{X}_{\mathcal{K}}; \hat{\mathbf{Y}}_{\mathcal{S}^c}\right)\right]^+, 
\end{equation}
for all $\mathcal{S} \subseteq \mathcal{L}$. Now, to show $R^*_{JD, SUM} \leq R^*_{SD, SUM}$, it suffices to show that each extreme point of $\left(R_{JD},\mathcal{P}_C\right)$ is dominated by a point in $\mathcal{P}^s_{SD}$ that achieves a sum rate greater or equal to the joint decoding sum rate $R_{JD}$.

To this end, define a set function $g: 2^{\mathcal{L}} \rightarrow \mathbb{R}$ as follows:
\begin{equation*}
g\left(\mathcal{S}\right):= R_{JD} + \sum_{\ell\in \mathcal{S}}I\left(\mathbf{Y}_{\ell};\hat{\mathbf{Y}}_{\ell}|\mathbf{X}_{\mathcal{K}}\right) - I\left(\mathbf{X}_{\mathcal{K}}; \hat{\mathbf{Y}}_{\mathcal{S}^c}\right),
\end{equation*}
for each $\mathcal{S} \subseteq \mathcal{L}$. It can be verified that the function $g^+\left(\mathcal{S}\right) = \max\left\{g\left(\mathcal{S}\right),0\right\}$ is a supermodular function (see Appendix~\ref{append:submodular}, Lemma~\ref{lem:supermodular-function}). By construction, $\mathcal{P}_C$ is equal to the set of $(C_1, R_2,\ldots, C_L)$ satisfying
\begin{equation*}
\displaystyle \sum_{\ell\in \mathcal{S}}C_{\ell}  \geq g^+\left(\mathcal{S}\right), \quad \forall \; \mathcal{S}\subseteq \mathcal{L}.
\end{equation*}

Following the results in submodular optimization (Appendix~\ref{append:submodular}, Proposition~\ref{prop:submodular-poly}), we have that for a  linear ordering $i_1 \prec i_2 \prec \cdots \prec i_K$ on the set $\mathcal{K}$, 
an extreme point of $\mathcal{P}_{C}$ can be computed as follows
\begin{equation*}
\tilde{C}_{i_j} =  g^+\left(\{i_1,\ldots, i_j\}\right) - g^+\left(\{i_1,\ldots, i_{j-1}\}\right).
\end{equation*}
All the $L!$ extreme points of $\mathcal{P}_{C}$ can be analyzed in the same manner. For notational simplicity we only consider the natural ordering $i_j = j$ in the following proof.

By construction,
\begin{multline*}
\tilde{C}_{j} = \left[ R_{JD} + \sum_{\ell=1}^jI\left(\mathbf{Y}_{\ell};\hat{\mathbf{Y}}_{\ell}|\mathbf{X}_{\mathcal{K}}\right)
- I\left(\mathbf{X}_{\mathcal{K}}; \hat{\mathbf{Y}}_{j+1}^L\right)\right]^+ \\
\quad - \left[ R_{JD} + \sum_{\ell=1}^{j-1}I\left(\mathbf{Y}_{\ell};\hat{\mathbf{Y}}_{\ell}|\mathbf{X}_{\mathcal{K}}\right) - I\left(\mathbf{X}_{\mathcal{K}}; \hat{\mathbf{Y}}_{j}^L\right)\right]^+.
\end{multline*}

Let $j$ be the first index for which $g\left(\{1,\ldots, j\}\right) > 0$. Then, by construction,
\begin{align*}
\tilde{C}_{k} =&  I\left(\mathbf{X}_{\mathcal{K}}; \hat{\mathbf{Y}}_{k}| \hat{\mathbf{Y}}_{k+1}^L\right) + I\left( \mathbf{Y}_{k}; \hat{\mathbf{Y}}_{k}| \mathbf{X}_{\mathcal{K}}\right) \\
 =& I\left(\mathbf{Y}_{k}; \hat{\mathbf{Y}}_{k}| \hat{\mathbf{Y}}_{k+1}^L\right)
\end{align*}
for all $k>j$, where the Markov chain $\hat{\mathbf{Y}}_{i} \leftrightarrow \mathbf{Y}_{i} \leftrightarrow
\mathbf{X}_{\mathcal{K}}\leftrightarrow \mathbf{Y}_j \leftrightarrow \hat{\mathbf{Y}}_j$, for $i\neq j$, is utilized in deriving the second equality. Clearly, all the $\tilde{C}_{k}$'s are in the successive decoding region $\mathcal{P}^s_{SD}$.

Moreover, we have $g\left(\{1,\ldots, j'\}\right) \leq 0$ for all $j' < j$. Thus, $\tilde{C}_{j}$ can be expressed as
\begin{eqnarray*}
\tilde{C}_{j} &=& R_{JD} + \sum_{\ell=1}^jI\left(\mathbf{Y}_{\ell};\hat{\mathbf{Y}}_{\ell}|\mathbf{X}_{\mathcal{K}}\right) - I\left(\mathbf{X}_{\mathcal{K}}; \hat{\mathbf{Y}}_{j+1}^L\right)\\
&=& \alpha I\left(\mathbf{Y}_{j+1}; \hat{\mathbf{Y}}_{j+1}| \hat{\mathbf{Y}}_{j+1}^L\right)
\end{eqnarray*}
where $\alpha \in [0,1]$ is defined as
\begin{equation*}
\alpha = \frac{R_{JD} + \sum\limits_{\ell=1}^jI\left(\mathbf{Y}_{\ell};\hat{\mathbf{Y}}_{\ell}|\mathbf{X}_{\mathcal{K}}\right) - I\left(\mathbf{X}_{\mathcal{K}}; \hat{\mathbf{Y}}_{j+1}^L\right)}{I\left(\mathbf{Y}_{j+1}; \hat{\mathbf{Y}}_{j+1}| \hat{\mathbf{Y}}_{j+1}^L\right)}.
\end{equation*}

Consider the two following successive decoding schemes:
\begin{itemize}
\item Scheme 1: The CP decodes quantization codewords $\hat{\mathbf{Y}}_{j+1},\ldots,  \hat{\mathbf{Y}}_{L}$ first, then decodes the message codewords $\mathbf{X}_{\mathcal{K}}$ sequentially. Note that the BSs with index $i\leq j$ are inactive, and are essentially removed from the network. The resulting extreme point $\mathbf{c}^{(1)} = (R^{(1)}_{SD}, C^{(1)}_1, \ldots, C^{(1)}_L)$ of $\mathcal{P}^s_{SD}$ satisfies
\begin{equation*}
\begin{cases}
\enspace C^{(1)}_i = 0, & \enspace \textrm{for} \enspace i\leq j,\\
\enspace C^{(1)}_i = I\left(\mathbf{Y}_{i};\hat{\mathbf{Y}}_{i}|\hat{\mathbf{Y}}^L_{i+1}\right) & \enspace \textrm{for} \enspace i> j, \\
\enspace R^{(1)}_{SD} = I\left(\mathbf{X}_{\mathcal{K}};\hat{\mathbf{Y}}_{j+1}^L\right).
\end{cases}
\end{equation*}
\item Scheme 2: The CP decodes quantization codewords $ \hat{\mathbf{Y}}_{j},\ldots,  \hat{\mathbf{Y}}_{L}$ first, then decodes the message codewords $\mathbf{X}_{\mathcal{K}}$ sequentially. Note that in this scheme, the BSs with index $i<j$ are inactive, and are essentially removed from the network. The resulting extreme point $\mathbf{c}^{(2)} = (R^{(2)}_{SD}, C^{(2)}_1, \ldots, C^{(2)}_L)$ of $\mathcal{P}^s_{SD}$ satisfies
\begin{equation*}
\begin{cases}
\enspace C^{(2)}_i = 0, & \enspace \textrm{for} \enspace i< j,\\
\enspace C^{(2)}_i = I\left(\mathbf{Y}_{i};\hat{\mathbf{Y}}_{i}|\hat{\mathbf{Y}}^L_{i+1}\right) & \enspace \textrm{for} \enspace i\geq j, \\
\enspace R^{(2)}_{SD} = I\left(\mathbf{X}_{\mathcal{K}};\hat{\mathbf{Y}}_{j}^L\right).
\end{cases}
\end{equation*}
\end{itemize}

Since $C_{\ell}$ is defined to be the maximum long-term average throughput of fronthaul link $\ell$,  the following point: $\mathbf{c}^{\alpha} = (1-\alpha) \mathbf{c}^{(1)} + \alpha \mathbf{c}^{(2)}$ lies in $\mathcal{P}^s_{SD}$. The corresponding sum rate $R_{SD}$ in $\mathbf{c}^{\alpha}$ is given by
\begin{align}
\label{eqn:timesharing-rate-SD}
& (1-\alpha)R^{(1)}_{SD} + \alpha R^{(2)}_{SD} \nonumber \\
& = (1-\alpha) I\left(\mathbf{X}_{\mathcal{K}}; \hat{\mathbf{Y}}_{j+1}^L\right) + \alpha I\left(\mathbf{X}_{\mathcal{K}}; \hat{\mathbf{Y}}_{j}^L\right) \nonumber\\
\displaystyle& \overset{(f)}{=} \frac{I\left(\mathbf{X}_{\mathcal{K}}; \hat{\mathbf{Y}}_{j}^L\right)-R_{JD} - \sum\limits_{\ell=1}^{j-1} I\left(\mathbf{Y}_{\ell};\hat{\mathbf{Y}}_{\ell}|\mathbf{X}_{\mathcal{K}}\right) }{I\left(\mathbf{Y}_{j+1}; \hat{\mathbf{Y}}_{j+1}| \hat{\mathbf{Y}}_{j+1}^L\right)} \nonumber \\
& \qquad \times I\left(\mathbf{X}_{\mathcal{K}}; \hat{\mathbf{Y}}_{j+1}^L\right) \nonumber \\
\displaystyle & \quad +  \frac{R_{JD} + \sum\limits_{\ell=1}^jI\left(\mathbf{Y}_{\ell};\hat{\mathbf{Y}}_{\ell}|\mathbf{X}_{\mathcal{K}}\right) - I\left(\mathbf{X}_{\mathcal{K}}; \hat{\mathbf{Y}}_{j+1}^L\right)}{I\left(\mathbf{Y}_{j+1}; \hat{\mathbf{Y}}_{j+1}| \hat{\mathbf{Y}}_{j+1}^L\right)} \nonumber \\
& \qquad \times I\left(\mathbf{X}_{\mathcal{K}}; \hat{\mathbf{Y}}_{j}^L\right) \nonumber \\
\displaystyle & \geq  \frac{R_{JD}\times\left[I\left(\mathbf{X}_{\mathcal{K}}; \hat{\mathbf{Y}}_{j}^L\right) - I\left(\mathbf{X}_{\mathcal{K}}; \hat{\mathbf{Y}}_{j+1}^L\right)\right]}{I\left(\mathbf{Y}_{j+1}; \hat{\mathbf{Y}}_{j+1}| \hat{\mathbf{Y}}_{j+1}^L\right)}  \nonumber \\
& \qquad + \frac{I\left(\mathbf{Y}_{j}; \hat{\mathbf{Y}}_{j}|\mathbf{X}_{\mathcal{K}}\right)\times I\left(\mathbf{X}_{\mathcal{K}}; \hat{\mathbf{Y}}_{j}^L\right)}{I\left(\mathbf{Y}_{j+1}; \hat{\mathbf{Y}}_{j+1}| \hat{\mathbf{Y}}_{j+1}^L\right)} \nonumber \\
\displaystyle & \overset{(g)}{\geq} R_{JD} \times \frac{I\left(\mathbf{X}_{\mathcal{K}}; \hat{\mathbf{Y}}_{j}^L\right) - I\left(\mathbf{X}_{\mathcal{K}}; \hat{\mathbf{Y}}_{j+1}^L\right) + I\left(\mathbf{Y}_{j}; \hat{\mathbf{Y}}_{j}|\mathbf{X}_{\mathcal{K}}\right)}{I\left(\mathbf{Y}_{j+1}; \hat{\mathbf{Y}}_{j+1}| \hat{\mathbf{Y}}_{j+1}^L\right)} \nonumber\\
& = R_{JD},
\end{align}
where the equality $(f)$ follows from the fact that $I\left(\mathbf{X}_{\mathcal{K}},\mathbf{Y}_{j+1}; \hat{\mathbf{Y}}_{j+1}| \hat{\mathbf{Y}}_{j+1}^L\right) = I\left(\mathbf{Y}_{j+1}; \hat{\mathbf{Y}}_{j+1}| \hat{\mathbf{Y}}_{j+1}^L\right)$, and inequality $(g)$ follows from the fact that $R_{JD} \leq I\left(\mathbf{X}_{\mathcal{K}}; \hat{\mathbf{Y}}_{j}^L\right)$.

Therefore, for every extreme point $(\tilde{C}_1,\ldots, \tilde{C}_L)$ of $\mathcal{P}_C$, the point $(R_{JD}, \tilde{C}_1,\ldots, \tilde{C}_L)$ is dominated by a point in $\mathcal{P}^s_{SD}$. This proves Theorem~\ref{thm:opt-VMAC}.

\section{Constant-gap Result for Compress-and-Forward with Joint Decoding}
\label{append:proof-JD-cons-gap}

The idea of the proof is to compare the
achievable rate of compress-and-forward with joint decoding with the following cut-set upper bound~\cite{El2011network}
\begin{multline}\label{eqn:cut-set-bound}
\sum_{k\in \mathcal{T}} R_k \leq \min \left\{\sum_{\ell\in \mathcal{S}} C_{\ell} \right. \\
 \left. + \log\frac{\left|\sum_{\ell\in \mathcal{S}^c}
\mathbf{H}^{\dagger}_{\ell,\mathcal{T}} \mathbf{\Sigma}_{\ell}^{-1} \mathbf{H}_{\ell,\mathcal{T}}  +
\mathbf{K}^{-1}_{\mathcal{T}}\right|}{\left|\mathbf{K}^{-1}_{\mathcal{T}}\right|}\right\}
\end{multline}
for all $\emptyset \subset \mathcal{T}\subseteq\mathcal{K}$ and $\mathcal{S}\subseteq\mathcal{L}$. In the expression of cut-set bound, the first term represents the cut across the fronthaul links in set $\mathcal{S}$, and the second term represents the cut from the users to the BSs in set $\mathcal{S}^c$.

Recall that the rate region for joint decoding (\ref{eqn:GaussRateRegion-JD}) under Gaussian quantization is the of $(R_1,\cdots,R_K)$ such that
\begin{multline*}
\sum_{k\in \mathcal{T}} R_k < \sum_{\ell\in \mathcal{S}}\left[C_{\ell} - \log \frac{|\mathbf{\Sigma}_{\ell}^{-1}|}{|\mathbf{\Sigma}_{\ell}^{-1} - \mathbf{B}_{\ell}|} \right] \\
 + \log\frac{\left|\sum_{\ell\in \mathcal{S}^c}
\mathbf{H}^{\dagger}_{\ell,\mathcal{T}} \mathbf{B}_{\ell} \mathbf{H}_{\ell,\mathcal{T}}  +
\mathbf{K}^{-1}_{\mathcal{T}}\right|}{\left|\mathbf{K}^{-1}_{\mathcal{T}}\right|}
\end{multline*}
for all $\emptyset \subset \mathcal{T}\subseteq\mathcal{K}$ and $\mathcal{S}\subseteq\mathcal{L}$, for some $0 \preceq \mathbf{B}_{\ell} \preceq \mathbf{\Sigma}^{-1}_{\ell}$.
We now show that if a rate tuple $(R_1,\cdots,R_K)$ is within the cut-set bound, then $(R_1-\eta,\cdots,R_K-\eta)$ is in the achievable rate region of joint decoding, where
\begin{multline}
|\mathcal{T}|\eta \leq \sum_{\ell\in \mathcal{S}}\log\frac{|\mathbf{\Sigma}_{\ell}^{-1}|}{|\mathbf{\Sigma}_{\ell}^{-1} - \mathbf{B}_{\ell}|} \\
+ \log\frac{\left|\sum_{\ell\in \mathcal{S}^c}
\mathbf{H}^{\dagger}_{\ell,\mathcal{T}} \mathbf{\Sigma}_{\ell}^{-1} \mathbf{H}_{\ell,\mathcal{T}}  +
\mathbf{K}^{-1}_{\mathcal{T}}\right|}{\left|\sum_{\ell\in \mathcal{S}^c}
\mathbf{H}^{\dagger}_{\ell,\mathcal{T}} \mathbf{B}_{\ell} \mathbf{H}_{\ell,\mathcal{T}}  +
\mathbf{K}^{-1}_{\mathcal{T}}\right|}
\end{multline}
is the gap between the cut-set bound and achievable rate of joint decoding.

Choose quantization noise level to be at the background noise level, i.e., $\mathbf{Q}_{\ell} = \mathbf{\Sigma}_{\ell}$. Then we have
\begin{equation*}
\mathbf{B}_{\ell} = (\mathbf{\Sigma}_{\ell} + \mathbf{Q}_{\ell})^{-1} = \frac{1}{2}\mathbf{\Sigma}^{-1}_{\ell}.
\end{equation*}
Evaluate gap $\eta$ with the above choice of $\mathbf{B}_{\ell}$ gives
\begin{equation*}
\eta \leq \frac{|\mathcal{S}|}{|\mathcal{T}|}\cdot N + M \leq NL + M,
\end{equation*}
which completes the proof of Proposition \ref{thm:constant-gap-JD}.

\bibliographystyle{IEEEtran}
\bibliography{IEEEabrv,yuhanthesis}

\begin{IEEEbiographynophoto}{Yuhan~Zhou}(S'08) received the B.E. degree in Electronic and Information Engineering from Jilin University, Jilin, China, in 2005, the M.A.Sc. degree from the University of Waterloo, ON, Canada, in 2009, and the Ph.D. degree from the University of Toronto, ON, Canada, in 2016, both in Electrical and Computer Engineering. Since 2016, he has been with Qualcomm Technologies Inc., San Diego, CA, USA. His research interests include wireless communications, network information theory, and convex optimization.
\end{IEEEbiographynophoto}

\begin{IEEEbiographynophoto}{Yinfei~Xu}(S'10) was born in July 1986 in Nanjing, China. He received the B.E. and Ph.D. degrees in 2008 and 2016, respectively,
both in Information Engineering, from Southeast University, Nanjing, China. Since March 2016, he has been in the Institute of Network Coding at The Chinese University of Hong Kong, Hong Kong, where he is currently a Postdoctoral Fellow. He was a visiting student in the Department of Electrical and Computer Engineering at McMaster University,  Hamilton, ON, Canada, from July 2014 to January 2015. His research interests include information theory, signal processing and wireless communications.
\end{IEEEbiographynophoto}

\begin{IEEEbiographynophoto}{Wei~Yu}(S'97-M'02-SM'08-F'14) received the B.A.Sc. degree in Computer Engineering and Mathematics from the University of Waterloo, Waterloo, Ontario, Canada in 1997 and M.S. and Ph.D. degrees in Electrical Engineering from Stanford University, Stanford, CA, in 1998 and 2002, respectively. Since 2002, he has been with the Electrical and Computer Engineering Department at the University of Toronto, Toronto, Ontario, Canada, where he is now Professor and holds a Canada Research Chair (Tier 1) in Information Theory and Wireless Communications. His main research interests include information theory, optimization, wireless communications and broadband access networks.

Prof. Wei Yu currently serves on the IEEE Information Theory Society
Board of Governors (2015-17). He is an IEEE Communications Society
Distinguished Lecturer (2015-16). He served as an Associate Editor for {\sc IEEE Transactions on Information Theory} (2010-2013), as an Editor for {\sc IEEE Transactions on Communications} (2009-2011), as an Editor for {\sc IEEE
Transactions on Wireless Communications} (2004-2007), and
as a Guest Editor for a number of special issues for the {\sc IEEE Journal on Selected Areas in
Communications} and the {\sc EURASIP Journal on Applied Signal Processing}. He was a Technical Program co-chair
of the IEEE Communication Theory Workshop in 2014, and a Technical
Program Committee co-chair of the Communication Theory Symposium at
the IEEE International Conference on Communications (ICC) in 2012. He
was a member of the Signal Processing for Communications and Networking
Technical Committee of the IEEE Signal Processing Society (2008-2013), then Vice Chair in 2016.
Prof. Wei Yu received a Steacie Memorial Fellowship in 2015, an IEEE
Communications Society Best Tutorial Paper Award in 2015, an IEEE ICC
Best Paper Award in 2013, an IEEE Signal Processing Society Best Paper
Award in 2008, the McCharles Prize for Early Career Research Distinction in
2008, the Early Career Teaching Award from the Faculty of Applied Science
and Engineering, University of Toronto in 2007, and an Early Researcher
Award from Ontario in 2006. He was named a Highly Cited Researcher by
Thomson Reuters in 2014 and 2015.
\end{IEEEbiographynophoto}

\begin{IEEEbiographynophoto}{Jun Chen}(S'03-M'06-SM'16) received the B.E. degree with honors in communication
engineering from Shanghai Jiao Tong University, Shanghai, China, in 2001
and the M.S. and Ph.D. degrees in electrical and computer engineering from
Cornell University, Ithaca, NY, in 2004 and 2006, respectively.

He was a Postdoctoral Research Associate in the Coordinated Science
Laboratory at the University of Illinois at Urbana-Champaign, Urbana, IL,
from September 2005 to July 2006, and a Postdoctoral Fellow at the IBM Thomas J. Watson
Research Center, Yorktown Heights, NY, from July 2006 to August 2007. Since September 2007 he has been with the Department of Electrical and Computer Engineering at McMaster University, Hamilton, ON, Canada, where he is currently an Associate Professor and a Joseph Ip Distinguished Engineering Fellow. His research interests include information theory,
wireless communications, and signal processing.

He received several awards for his research, including the Josef Raviv
Memorial Postdoctoral Fellowship in 2006, the Early Researcher Award from
the Province of Ontario in 2010, and the IBM Faculty Award in 2010. He is currently serving as an Associate Editor for Shannon Theory for the IEEE Transactions on Information Theory.
\end{IEEEbiographynophoto}

\end{document}